# The fate of compound with AgF$_2$ : AgO stoichiometry – a theoretical study

Mateusz Domański and Wojciech Grochala*
Center of New Technologies, University of Warsaw, Żwirki i Wigury 93, 02089 Warsaw
Poland

* w.grochala@cent.uw.edu.pl

This work is dedicated to Leszek Stolarczyk at his 70$^{th}$ birthday

**Abstract**
Metal oxyfluorides constitute a broad group of chemical compounds with rich spectrum of crystal structures and properties. Here we predict, based on evolutionary algorithm approach, the crystal structure and selected properties of Ag$_2$OF$_2$. This system may be considered as the 1:1 'adduct' of AgF$_2$ (i.e. an antiferromagnetic charge transfer positive U insulator) and AgO (i.e. a disproportionated negative U insulator). We analyze oxidation states of silver in each structure, possible magnetic interactions, as well as energetic stability. Prospect is outlined for synthesis of polytypes of interest using diverse synthetic approaches.

**Introduction**
Silver tends to occur in nature either as elemental metal (i.e. native silver) or in ores; in all of the latter it shows the monovalent (I) formal oxidation state. Chemists can push oxidation states of silver to higher values (II, III, V) among which divalent state is represented by the largest number of compounds (mostly fluorides) in excess of 100.[1,2,3] Many of these are of interest to solid state physics and chemistry due to their unusually strong magnetic superexchange,[4,5] as well as possibility to generate superconductivity in quasi-2D systems.[1,6] AgF$_2$ is a prototypical compound of divalent silver; it features paramagnetic Ag(II) sites with d$^9$ electronic configuration of transition metal cation, and adopts a puckered-sheet layered structure with an elongated octahedral coordination of Ag (Figure 1a).[7,8] Nearly all silver(II) fluorides recite the same structural features i.e. distorted octahedral coordination of Ag(II) and the presence of genuine divalent silver site.[1] An oxide analogue of AgF$_2$, i.e. AgO, first prepared in 1954, has been initially believed to contain Ag(II).[9,10] Subsequent studies have shown that it is in fact a mixed valent compound with frozen Ag(I) and Ag(III) valencies.[11,12] The compound is diamagnetic, consistent with the presence of diamagnetic Ag1+ (d$^{10}$ electronic configuration) and low-spin Ag3+ (d$^8$ configuration) (Figure 1b). In 1986 second polymorphic form of AgO was prepared;[13] its tetragonal form i salso diamagnetic and mixed-valent system, just like its previously known monoclinic analogue.

Metal oxyfluorides constitute a broad group of chemical compounds with rich spectrum of crystal structures and properties.[14,15] Since both AgF$_2$ and AgO adopt distorted fcc lattices, it might therefore be anticipated that formation of a sort of intergrowth structure or mixed-ligand (oxofluoride) one should be facile. However, theoretical studies of electronic structure of AgF$_2$[1,6] and AgO[16,17] as well as XPS studies of both compounds,[18,19,20,21] have indicated substantial covalence of metal-ligand bonding. However, there is a marked difference between them: in the Mott-Hubbard picture, AgF$_2$ is a typical 'positive U' system, while AgO is a rare 'negative U' one.[1,6] This has to do with the fact that Ag(II) is a powerful oxidizer and its 4d states are found at so large binding energies, that they are able to introduce holes to the O(2p) states.

Interestingly, the metal oxidation states in both compounds are preserved at elevated pressure. $AgF_2$ retains paramagnetic Ag(II) up to 40 GPa,[22] while AgO preserves mized-valence up to at least 80 GPa[23] (and in striking contrast to its copper analogue). According to theoretical calculations, such chemical chatacteristics should pertain to at least 100 GPa, with both compounds showing a finite band gap at the Fermi level.[20,21]

There is one more noticeable difference between $AgF_2$ and AgO; the former is stable with respect to thermal decomposition to AgF and ½ $F_2$, while the latter is unstable with respect to ½ $Ag_2O$ and ¼ $O_2$. This is yet another exemplification of the entirely different electronic structure and different share of ligand (nonmetal) states to chemical bonding close to the Fermi level of both compounds.

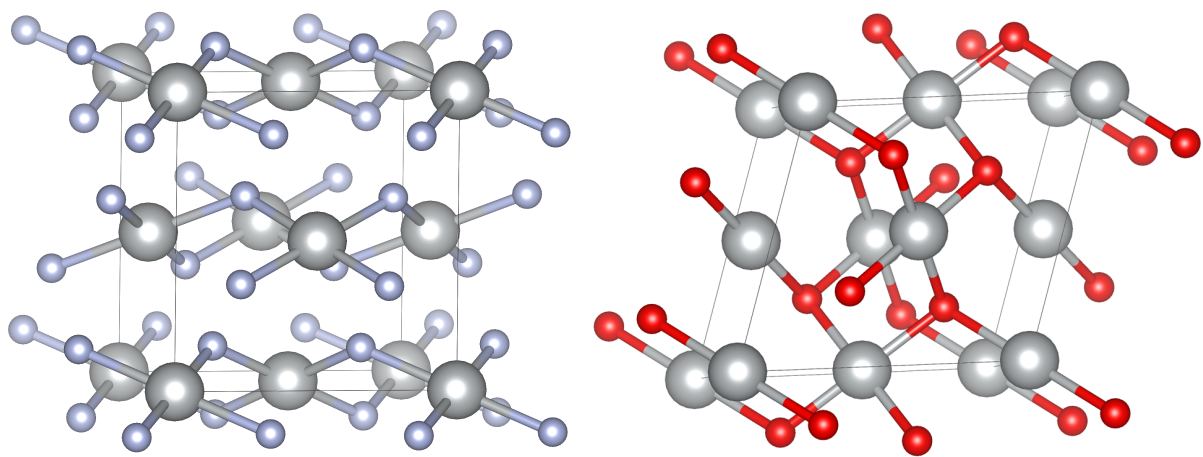

Figure 1. Crystal structure and local coordination polyhedra around Ag for: left – $AgF_2$, right – monoclinic AgO.

Given that $AgF_2$ and AgO, unlike vast majority of metal fluorides and related oxides, show extremely different physicochemical features, and they belong to different classes of compounds in Robin and Day characteristics,[24] as well as from the viewpoint of Zaanen-Sawatzky-Allen diagram,[25] it is justified to ask what structure and properties would be exhibited by stoichiometry which contains half of both, *i.e.* $AgF_2$ : AgO 1:1 'adduct'. Would it contain Ag(I), Ag(II), and Ag(III) cations simultaneously, or would it exhibit electronic character typical only of one of its parent compounds? Would solely Ag(II) be present yielding very strong spin polarization of ligands (particularly oxide ones) with concomitant resulting bulk magnetic ordering, or would $Ag_2OF_2$ be disproportionated and diamagnetic, just like AgO? Would it be stable with respect to loss of ligands and decrease of the formal oxidation state of a metal, or not?

In this work we look at this problem using quantum mechanical calculations in conjunction with the evolutionary algorithm approach. The density functional theory (DFT) calculations, as we will see, lead to a rather unexpected outcome.

**Theoretical calculations**

The quest for lowest-energy crystal structures was conducted using XtalOpt[26,27] (which utilizes particle-swarm evolutionary algorithm) in combination with VASP and using rotationally invariant DFT+U method proposed by Liechtenstein.[28] We applied GGA (general gradient approximation) with Perdew-Burke-Ernzerhof functional adapted for solids (PBEsol)[29]. This particular DFT+U approach was chosen because of its success in the solid state physics and chemistry studies. We used value of U and J (Coulomb and exchange parameters) equal 5.0 and 1.0 eV only for silver atoms, according to Kasinathan et al.[30] (and later used many times in numerous computational studies). The projector–augmented-wave (PAW) method was used,[31] as implemented in the VASP 5.4.4 program.[32] The plane waves cut-off energy was set to 520 eV with a self-consistent-field convergence criterion of $10^{-7}$ eV. Valence electrons were treated explicitly, while recommended VASP pseudopotentials[33] (providing scalar relativistic effects) were used for the description of core electrons. Ionic relaxation was performed using a conjugate-gradient algorithm until the energy difference was below $10^{-5}$ eV threshold. In the finest optimization step (at the end) k-point mesh was set at 2π x 0.05 Å$^{-1}$. Energy tresholds in XtalOpt runs were 5 times higher (i.e. 5·$10^{-7}$ eV for electronic and 5·$10^{-5}$ eV for ionic convergence) and the densest k-point mesh was set to 2π x 0.10 Å$^{-1}$. Electronic density of states was calculated using above-mentioned DFT+U approach as well as SCAN meta-GGA functional and HSE06 hybrid functional, as implemented in VASP software.

Using the evolutionary algorithms, we have considered unit cells containing 4 formula units and generated a pool of 600 structures, including 526 unique ones (for cif files *cf.* Appendix 1). Crystal structures based on $Sn_2OF_2$,[34] and two $Pb_2OF_2$ polytypes,[35,36] have been also calculated. The same procedure was applied to the structure of $Ag_2O$ in which additional F atom has been added at (¼,¼,¾) position yielding $Ag_2OF_3$ formula (Z=4); subsequently, 1/3 of F atoms were removed generating four alternative patterns of vacancies. Initially, these structures were optimized while preserving their original symmetry; in subsequent step symmetry constraints were released, and the resulting optimized cells were symmetrized again. Only the lowest-energy polytypes are presented here.

Spin polarized calculations were carried out for all structures obtained from XtalOpt and other proposed crystal structures. For the structures featuring uncompensated spin on metal and/or nonmetal atoms, diferent magnetic models were constructed to find the most stable one. Only the lowest energy models are listed in this work. Spin polarized calculations were carried out using broken-symmetry method for the most interesting case.[37, 38,39]
Aside from $Ag_2OF_2$, calculations were carried out for diamagnetic $Ag_2O$, antiferromagnetic $AgF_2$, monoclinic AgO, AgF, antiferromagnetic α-$O_2$, as well as α-$F_2$, as potential substrates or products of chemical reactions in the ternary Ag-O-F phase diagram.

All visualizations of the structures were performed using VESTA software.[40]

**Results and analysis**

Extremely few metal oxofluorides of the formula $M_2OF_2$ have been prepared so far. These consist of monoclinic $Sn_2OF_2$ as well as two tetragonal forms of its Pb analogue (Figure 2).[32,33,34] In $Sn_2OF_2$ two Sn(II) sites are found: one with butterfly coordination (bond lengths Sn-F 2x 2.388 Å, Sn-O 2x 2.106 Å) and one in distorted trigonal pyramidal coordination (Sn-F 2x 2.139 Å, Sn-O 1x 2.036 Å). Both coordination types are consequence of stereochemical activity of the lone pair at Sn(II) site.

In $Pb_2OF_2$ (*P*–4m2 form), all Pb(II) sites exhibit ireegular coordination sphere with two Pb-O bonds at 2.14 Å and 2.54 Å, and four Pb-F contacts (two at 2.512 Å, one at 2.628 Å, and one at 2.6285 Å). The same compound in its $P4_2/nmc$ form shows two inequivalent Pb sites; one binds to two oxygen atoms in trans geometry (2.29 Å, 2.52 Å) and to four F atoms (2x 2.332 Å, 2x 2.70 Å), and another to two oxygen atoms in the cis geometry (2.12 Å, 2.69 Å), and to four F atoms (2x 2.332 Å, 2x 2.70 Å). Again, stereochemical activity of the lone pair at Pb is reflected in deformation of the coordination olyhedra from their hypothetical higher symmetry forms. Worth noting is a fact, that rarely oxyfluorides possess cations on the oxidation state lower than the highest-possible; thus, structures of both Sn and Pb oxyfluorides are such exceptions.

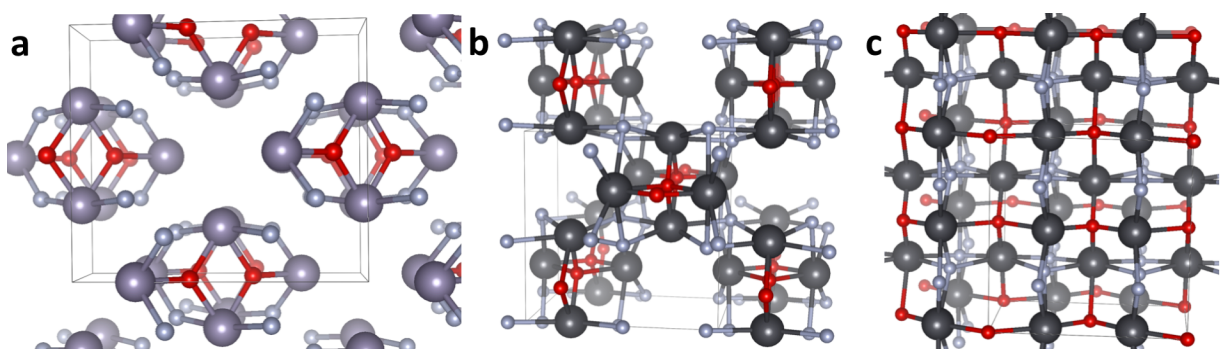

Figure 2. Crystal structure of a. $Sn_2OF_2$ (*C*2/m), b. $Pb_2OF_2$ (*P*–4m2), c. $Pb_2OF_2$ (*P*$4_2$/nmc). Sn – mauve, Pb – black, O – red, F – light blue balls.

None of the above-mentioned structure types hosts local geometry of metal which is typical for Ag; therefore, substantial structural changes are expected upon optimization of structures with Sn or Pb substituted for Ag. We will return to those below.

In the pool of 526 unique structures generated by Xtalopt and optimized by VASP, and supplemented by three structures obtained from Sn and Pb analogues, and one coming from F addition to $Ag_2O$ structure, twenty-two overall attracted our special attention due to their unique structural features (Table 1). We label structures originating from XtalOpt in order of increasing energy. Other structures are identified by their parent polytype.

Table 1. Energy (eV) and volume (Å$^3$) of obtained structures, and their differences with respect to lowest-energy product, $O_2 + 2·AgF$, as well as to metastable binary substrates, $AgO + AgF_2$, and to $Ag_2O + F_2$. Structures originating from XtalOpt quest are labeled by their numer in the generated pool, while those coming from modifications of the known structures are labeled by the chemical formula and polytype. Structures are ranked with respect to their computed energy.

| Label | E / Z | V / Z | ΔE / Z | ΔV / Z | ΔE / Z | ΔV / Z | ΔE / Z | ΔV / Z |
|---|---|---|---|---|---|---|---|---|
| | | | (vs $O_2 + 2·AgF$) | | (vs $AgO + AgF_2$) | | (vs $Ag_2O + F_2$) | |
| 1 | -17.038 | 72.6 | 0.215 | 3.2 | -0.2734 | 6.3 | -3.6995 | -12.9 |
| 2 | -17.036 | 68.1 | 0.218 | -1.3 | -0.2710 | 1.8 | -3.6835 | -13.8 |
| 3 | -17.013 | 66.7 | 0.240 | -2.7 | -0.2487 | 0.4 | -3.6612 | -15.2 |
| 4 | -17.012 | 73.2 | 0.241 | 3.8 | -0.2479 | 6.9 | -3.6603 | -8.7 |
| 5 | -17.005 | 66.6 | 0.248 | -2.8 | -0.2407 | 0.4 | -3.6531 | -15.2 |
| 6 | -17.004 | 66.7 | 0.249 | -2.7 | -0.2394 | 0.4 | -3.6518 | -15.2 |
| 7 | -17.000 | 71.6 | 0.253 | 2.2 | -0.2353 | 5.3 | -3.6478 | -10.3 |
| 8 | -16.993 | 68.7 | 0.260 | -0.6 | -0.2283 | 2.5 | -3.6407 | -13.1 |
| 9 | -16.990 | 73.3 | 0.263 | 4.0 | -0.2251 | 7.1 | -3.6375 | -8.5 |
| 10 | -16.978 | 69.2 | 0.275 | -0.2 | -0.2131 | 2.9 | -3.6256 | -12.7 |
| 12 | -16.964 | 65.4 | 0.289 | -4.0 | -0.1991 | -0.9 | -3.6115 | -16.5 |
| 16 | -16.956 | 68.0 | 0.297 | -1.3 | -0.1914 | 1.8 | -3.6038 | -13.8 |
| 62 | -16.784 | 70.8 | 0.469 | 1.4 | -0.0192 | 4.6 | -3.4317 | -11.0 |
| 171 | -16.632 | 67.0 | 0.621 | -2.4 | 0.132 | 0.7 | -3.280 | -14.9 |
| 219 | -16.564 | 73.1 | 0.689 | 3.7 | 0.2007 | 6.8 | -3.2117 | -8.8 |
| Ag$_2$O + 2·F in vacancies (P1) | -16.375 | 68.8 | 0.878 | -0.6 | 0.390 | 2.5 | -3.023 | -13.1 |
| Pb$_2$OF$_2$ type (Pnnm) | -16.368 | 70.3 | 0.885 | 0.9 | 0.396 | 4.0 | -3.016 | -11.6 |
| Sn$_2$OF$_2$ type (C2/m) | -16.270 | 69.2 | 0.983 | -0.1 | 0.495 | 3.0 | -2.918 | -12.6 |
| 346 | -16.258 | 95.3 | 0.995 | 25.9 | 0.5065 | 29.0 | -2.9060 | 13.5 |
| Pb$_2$OF$_2$ type (P-4m2) | -16.123 | 68.3 | 1.130 | -1.1 | 0.641 | 2.1 | -2.771 | -13.5 |
| 423 | -15.812 | 66.9 | 1.440 | -2.4 | 0.9518 | 0.7 | -2.4606 | -14.9 |
| 461 | -15.132 | 70.0 | 2.121 | 0.6 | 1.632 | 3.7 | -1.780 | -11.9 |

### I. Structures featuring O$_2$ units.

All the lowest-energy structures obtained with XtalOpt algorithm (i.e. ten lowest-energy ones and more of higher energies) contain O atoms in form of distinct O-O bonded O$_2$ units (Figure 3). The first structure containing other oxygen form (i.e. both isolated O centers together with O$_2$ units) is only 62$^{th}$ in energy rank. The O$_2$ units exhibit the computed bond length varying from 1.233 Å (in structure No.2) up to 1.293 Å (in structure No.4), with intermediate values for remaining ten structures (Table 2). Such O-O bond length is close to those for elemental O$_2$ (1.208 Å in ground triplet state $^3\Sigma^-_g$, 1.220 Å in first excited singlet state $^1\Delta_g$) or O$_2^{-•}$ anion-radical (1.28 Å). Simultaneously, in majority of these structures there is substantial spin density on each O atom (Table 3). For example, for structure No.1, magnetic moment on O1 atom is +0.43 $\mu_B$, and +0.42 $\mu_B$ on O4 (which is bound to O1). Corresponding spin densities on O2 and O3 atoms (also bound in a single molecule) are negative of those for O1 and O4, respectively. Therefore, it is clear that both O$_2$ species present in structure No.1 are similar to triplet O$_2$ molecules, albeit with a smaller total magnetic moment than expected. Very similar situation is found for structures No. 2, 3, 5, 6, 7, and 10. On the other hand, structures No.4, 9 and 12 feature spin densities on all O atoms which are either null, or residual denisites, which are very small (their absolute values not exceeding 0.03 $\mu_B$). This, together with slightly longer O-O bond lengths in these structures, indicates the presence of species similar to singlet O$_2$. The lowest energy structure of this kind, No.4, has energy which is merely 0.026 eV higher per Ag$_2$OF$_2$ formula (hence ca. 0.052 eV higher per O$_2$ molecule), than the lowest energy structure featuring triplet O$_2$ molecules (No.1). This result may be surprisng given that the lowest energy singlet state for isolated O$_2$ molecule in the gas phase is as much as 0.9773 eV above the ground state triplet. However, it must be remembered that O$_2$ molecules in the crystal structures No.1-16 are not isolated, but they strongly intract with Ag sites (as it will be discussed below). These interactions may also reduce magnetic moment in these structures which contains quasi-triplet O$_2$ species. Note: magnetic moment on F atoms is usually very small in all structures.

Table 2. The O-O and Ag-O bond lengths (in Å) as indicated for the 1st and the 2nd molecule of $O_2$ present in the crystal structures. Note, for structure No.62 only one crystallographically independent molecule is present.

| Rank | d(O-O) 1st mol. | d(Ag-O) 1st mol. | d(O-O) 2nd mol. | d(Ag-O) 2nd mol. | d($O_2$-$O_2$) |
|---|---|---|---|---|---|
| 1 | 1.239 | 2.322 | 1.239 | 2.325 | 2.073 |
| 2 | 1.239 | 2.326 | 1.233 | 2.394 | 2.024 |
| 3 | 1.256 | 2.280 | 1.256 | 2.280 | 2.059 |
| 4 | 1.274 | 2.278 | 1.293 | 2.284 | 2.730 |
| 5 | 1.254 | 2.333 | 1.254 | 2.329 | 2.197 |
| 6 | 1.268 | 2.365 | 1.245 | 2.407 | 2.067 |
| 7 | 1.277 | 2.197 | 1.284 | 2.208 | 2.672 |
| 8 | 1.245 | 2.401 | 1.245 | 2.277 | 1.891 |
| 9 | 1.276 | 2.183 | 1.270 | 2.321 | 3.619 |
| 10 | 1.273 | 2.248 | 1.289 | 2.187 | 2.522 |
| 12 | 1.249 | 2.312 | 1.253 | 2.465 | 1.975 |
| 16 | 1.273 | 2.093 | 1.282 | 2.195 | 2.743 |
| 62 | 1.248 | 2.319 | - | - | 5.687 |

Table 3. Magnetic moments on atoms (in $\mu_B$) for all considered structures. S(tot) stands for total magnetic moment per unit cell.

| Rank | S(tot) | Ag1 | Ag2 | Ag3 | Ag4 | Ag5 | Ag6 | Ag7 | Ag8 | O1 | O2 | O3 | O4 | F1 | F2 | F3 | F4 | F5 | F6 | F7 | F8 |
|---|---|---|---|---|---|---|---|---|---|---|---|---|---|---|---|---|---|---|---|---|---|
| 1 | 0.0 | 0.00 | 0.00 | 0.10 | 0.00 | -0.10 | 0.00 | 0.00 | 0.00 | 0.43 | -0.43 | -0.42 | 0.42 | 0.01 | 0.00 | 0.00 | 0.00 | 0.00 | -0.01 | 0.00 | 0.00 |
| 2 | 0.0 | 0.00 | 0.03 | 0.08 | 0.03 | -0.01 | -0.07 | 0.00 | -0.02 | 0.42 | -0.44 | 0.41 | -0.43 | 0.02 | -0.01 | 0.00 | 0.00 | 0.01 | -0.01 | 0.00 | 0.00 |
| 3 | 0.0 | 0.06 | -0.06 | 0.00 | 0.06 | -0.01 | -0.06 | 0.01 | 0.00 | -0.31 | 0.31 | -0.31 | 0.31 | 0.00 | 0.00 | 0.01 | -0.01 | 0.00 | 0.00 | 0.00 | 0.00 |
| 4 | -0.2 | 0.11 | 0.07 | 0.01 | -0.13 | -0.05 | -0.10 | 0.11 | -0.15 | -0.03 | 0.02 | -0.02 | 0.03 | 0.01 | -0.02 | 0.01 | -0.01 | 0.00 | -0.02 | 0.00 | 0.00 |
| 5 | 0.0 | 0.03 | 0.05 | -0.06 | 0.06 | 0.00 | -0.05 | 0.00 | -0.03 | 0.46 | -0.45 | 0.45 | -0.46 | 0.00 | -0.01 | 0.01 | -0.01 | 0.00 | 0.00 | 0.01 | 0.00 |
| 6 | 0.0 | -0.04 | 0.07 | 0.03 | -0.08 | 0.01 | 0.02 | -0.07 | -0.08 | -0.31 | 0.39 | -0.32 | 0.39 | -0.01 | -0.03 | -0.02 | 0.00 | 0.00 | -0.01 | 0.02 | -0.02 |
| 7 | 0.0 | -0.07 | 0.16 | 0.00 | 0.01 | 0.00 | -0.11 | 0.12 | -0.15 | 0.52 | -0.57 | 0.62 | -0.53 | -0.03 | -0.01 | 0.00 | 0.01 | 0.02 | 0.02 | 0.00 | -0.02 |
| 8 | 1.4 | 0.00 | 0.00 | 0.00 | 0.00 | 0.00 | 0.00 | 0.00 | 0.00 | 0.75 | 0.19 | 0.60 | -0.14 | 0.00 | 0.00 | 0.00 | 0.00 | 0.00 | 0.00 | 0.00 | 0.00 |
| 9 | 0.0 | 0.00 | 0.09 | 0.01 | 0.18 | -0.02 | -0.11 | -0.08 | -0.03 | -0.03 | -0.02 | 0.03 | -0.01 | -0.01 | 0.00 | 0.02 | 0.00 | -0.02 | 0.01 | 0.01 | 0.00 |
| 10 | 0.0 | 0.13 | -0.02 | -0.07 | 0.15 | 0.04 | 0.05 | -0.10 | -0.07 | -0.58 | -0.60 | 0.55 | 0.51 | 0.00 | -0.01 | 0.02 | -0.01 | 0.00 | 0.01 | 0.00 | 0.03 |
| 12 | 0.0 | 0.00 | 0.00 | 0.00 | 0.00 | 0.00 | 0.00 | 0.00 | 0.00 | 0.00 | 0.00 | 0.00 | 0.00 | 0.00 | 0.00 | 0.00 | 0.00 | 0.00 | 0.00 | 0.00 | 0.00 |
| 16 | -1.0 | -0.03 | 0.00 | -0.06 | 0.01 | 0.04 | 0.04 | 0.05 | 0.02 | -0.67 | 0.00 | -0.70 | 0.28 | -0.01 | 0.00 | 0.00 | 0.01 | 0.01 | 0.01 | 0.00 | 0.01 |
| 62 | -0.2 | 0.03 | 0.00 | 0.40 | 0.04 | -0.47 | 0.08 | -0.48 | -0.42 | -0.34 | -0.21 | 0.69 | 0.73 | -0.06 | 0.06 | -0.08 | -0.07 | 0.08 | -0.05 | -0.07 | -0.06 |
| 171 | -0.9 | -0.34 | 0.47 | 0.02 | 0.01 | -0.04 | -0.50 | -0.42 | -0.24 | 0.04 | -0.05 | 0.22 | 0.14 | -0.14 | -0.02 | 0.06 | 0.05 | 0.05 | -0.12 | -0.02 | -0.12 |
| 219 | 0.0 | 0.37 | 0.39 | -0.34 | 0.45 | -0.42 | -0.38 | 0.42 | -0.39 | -0.03 | 0.01 | 0.03 | -0.17 | 0.04 | -0.01 | 0.03 | 0.01 | -0.01 | 0.01 | 0.00 | 0.03 |
| 346 | -0.8 | 0.36 | 0.36 | 0.36 | 0.36 | -0.36 | -0.36 | -0.36 | -0.36 | -0.30 | -0.15 | 0.05 | -0.41 | 0.11 | 0.11 | 0.11 | 0.11 | -0.11 | -0.11 | -0.11 | -0.11 |
| 423 | 0.0 | 0.00 | 0.00 | 0.00 | 0.00 | 0.00 | 0.00 | 0.00 | 0.00 | 0.00 | 0.00 | 0.00 | 0.00 | 0.00 | 0.00 | 0.00 | 0.00 | 0.00 | 0.00 | 0.00 | 0.00 |
| 461 | 0.7 | 0.21 | 0.25 | 0.21 | 0.25 | -0.21 | -0.25 | -0.21 | -0.25 | 0.66 | 0.00 | 0.00 | 0.00 | 0.00 | 0.00 | 0.00 | 0.00 | 0.00 | 0.00 | 0.00 | 0.00 |
| $Pb_2OF_2$ (P-4m2) | -1.9 | 0.04 | 0.02 | 0.02 | 0.04 | -0.24 | -0.25 | -0.24 | -0.23 | 0.06 | -0.78 | -0.08 | -0.10 | -0.01 | -0.01 | -0.01 | -0.01 | -0.02 | -0.03 | -0.02 | -0.02 |
| $Pb_2OF_2$ (Pnnm) | 0.0 | 0.00 | 0.35 | 0.00 | 0.35 | -0.35 | -0.35 | 0.00 | 0.00 | 0.00 | 0.00 | 0.00 | 0.00 | 0.00 | 0.00 | 0.00 | 0.00 | 0.00 | 0.00 | 0.00 | 0.00 |
| $Sn_2OF_2$ (C2/m) | 7.6 | 0.48 | 0.48 | 0.48 | 0.48 | 0.55 | 0.55 | 0.55 | 0.55 | 0.63 | 0.63 | 0.63 | 0.63 | 0.12 | 0.12 | 0.12 | 0.12 | 0.12 | 0.12 | 0.12 | 0.12 |
| $Ag_2O$ + 2·F (P1) | 0.0 | 0.07 | -0.31 | 0.00 | 0.26 | - | - | - | - | -0.01 | -0.02 | - | - | 0.04 | 0.04 | -0.03 | -0.03 | - | - | - | - |

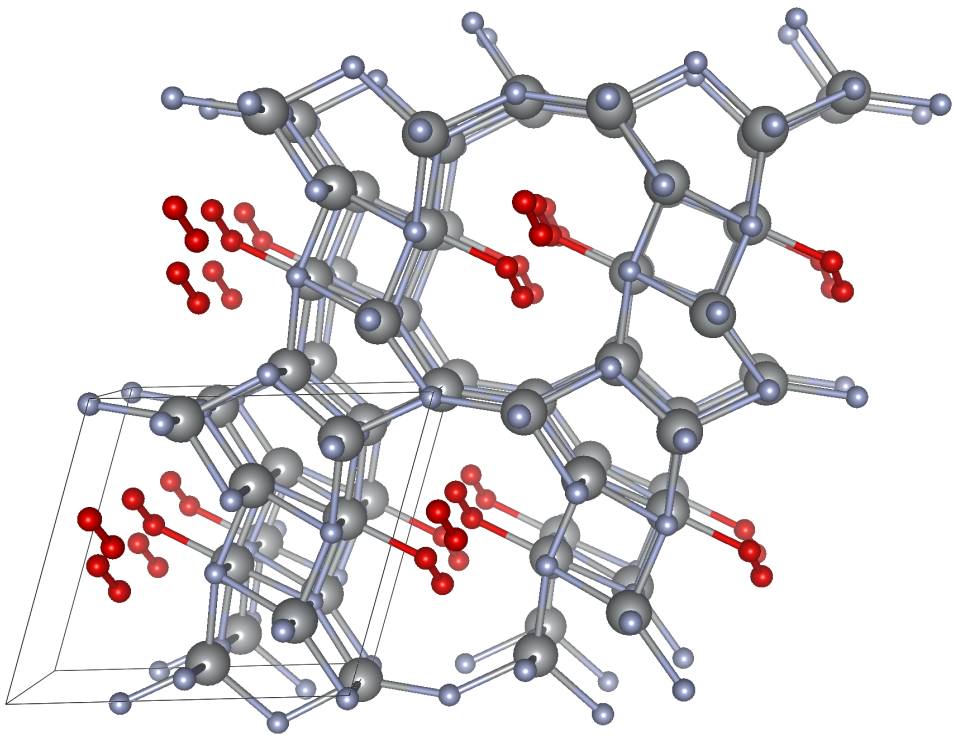

Figure 3. Crystal structure of polytype No.1. Ag – gray, O – red, F – light blue balls.

Structure No.8 is different from those described above since it features $O_2$ species, with very different spins on each of two $O_2$ subunits. N particular, spin densities on O1 and O3 atoms (bound by a chemical bond) are +0.75 $\mu_B$ and +0.60 $\mu_B$, respectively. Simultaneously, magnetic moments on O2 and O4 atoms are much smaller, +0.19 $\mu_B$ and –0.14 $\mu_B$, respectively. This suggests the presence of both triplet (O2-O3) and singlet (O1-O4) molecules in the structure, and some spin contamination in the latter is natural consequence of its proximity to the former. As a consequence, there is a net (uncompensated) magnetic moment in the unit cell.

The case of structure No.16 is still different, since more asymmetric spin density within each $O_2$ subunit. Notably, spin on O2 atom is null, while that on O3 atom (bound to O2) is –0.70 $\mu_B$. Similarly, spins on O1 and O4 atoms (bound together) are –0.67 $\mu_B$ and +0.28 $\mu_B$. At the first sight, such feature might imply the presence of quasi-$O_2^{-\bullet}$ anion-radical in the structures. However, the formulation of $Ag_4O_2F_4$ with $O_2^{-\bullet}$ anion-radical would be associated with the formal oxidation state of Ag of Ag(II) on one of four Ag centers. Inspection of Table 3 shows no appreciable spin density on any Ag site in this structure, hence Ag(II) is absent here. Therefore, we conclude that these $O_2$ species correspond formally to neutral $O_2$ units, and asymmetry of spin density within these units is caused by different coordination of both terminals to diverse Ag sites (more on that anon).

All discussed polymorphs feature $O_2$ molecules contained within voids in the crystal structure; sometimes channels filled with $O_2$ molecules are formed. Usually $O_2$ species coordinate to some coordinateively unsaturated Ag(I) site(s), either in terminal or bridging fashion. This is due to donor acceptor and electrostatic interactions between moieties involved in chemical bonding. The resulting polymorphs resemble other known clathrate structures such as Xe/$H_2O$, $Ba_xSi_{46}$, etc. The topology of [$Ag_2F_2$] cages is of interest (Figure 4).

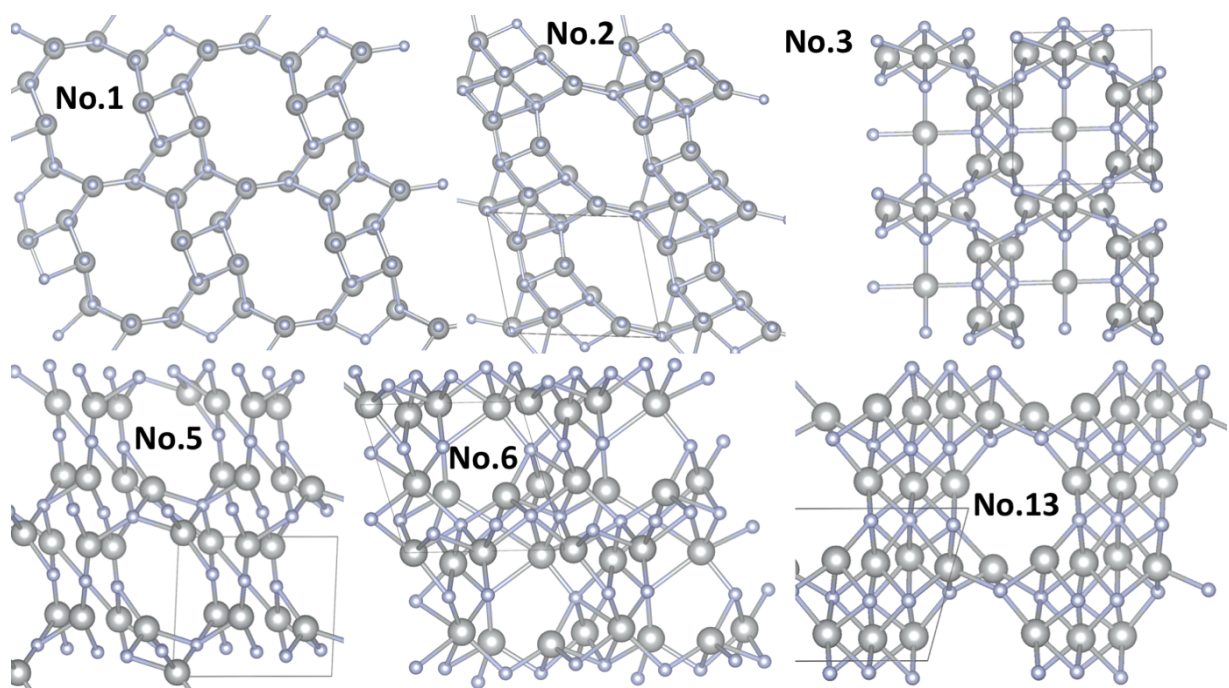

Figure 4. Selected projections of [Ag$_2$F$_2$] sublattice of polytypes No.1, 2, 3, 5, 6 and 13. Ag – gray, O – red, F – light blue balls.

A huge richness of [Ag$_2$F$_2$] sublattices at rather small energy differences between different structural types might suggest existence of very rich polymorphism of Ag$_2$F$_2$O featuring O$_2$ molecules. However, it could be that all of them are artifactual since they originate from finite numer of molecules per unit cell taken into calculations, as well as boundary periodic conditions. It is uncertain whether any of those could be prepared in experiment. Note that the energy of structure No.1, most stable of those, is still 0.215 eV above the mixture of 2AgF and O$_2$. This is indicative of a facile phase separation. O$_2$ is known to be a very weak Lewis base, and its bonding to Ag(I) (here typically at 2.1–2.4 Å) cannot compensate for the disrupture of ionic network of bulk AgF. Indeed, only small numer of O$_2$ complexes has been prepared but these are usually superoxo or peroxo species.[41,42]

II. **Structure featuring O$_2$ units and isolated O ligands.**

Structure No.62 also features the presence of O$_2$ subunit (with the bond length of 1.248 Å between O3 and O4 atoms), but there is only one such species per formula Ag$_8$O$_4$F$_8$ (Figure 5). There is substantial spin on both O atoms (0.69–0.73 $\mu_B$) as typical for triplet O$_2$ molecule. This suggests the ionic formulation as Ag(I)$_4$Ag(II)$_4$(O$_2$)(O$^{-2}$)$_2$(F$^-$)$_8$. Indeed, there is appreciable spin density on half of Ag sites, namely Ag3 (0.40 $\mu_B$), Ag5 (0.40 $\mu_B$), Ag7 (–0.48 $\mu_B$), and Ag8 (–0.42 $\mu_B$), which confirms the presence of divalent silver. This unusual mixed-valence polymorph has energy of 0.254 eV above our ground state structure (per formula unit Ag$_2$OF$_2$). Structure No.62 contains not only O$_2$ units but also isolated O centers, and as such it is intermediate between structures containing O$_2$ molecules and those which do not show their presence. Given that only half of O atoms is unpaired in molecules, approximate energy penalty of bond breaking in O$_2$ units found in these structures is some 1 eV per O$_2$ molecule.

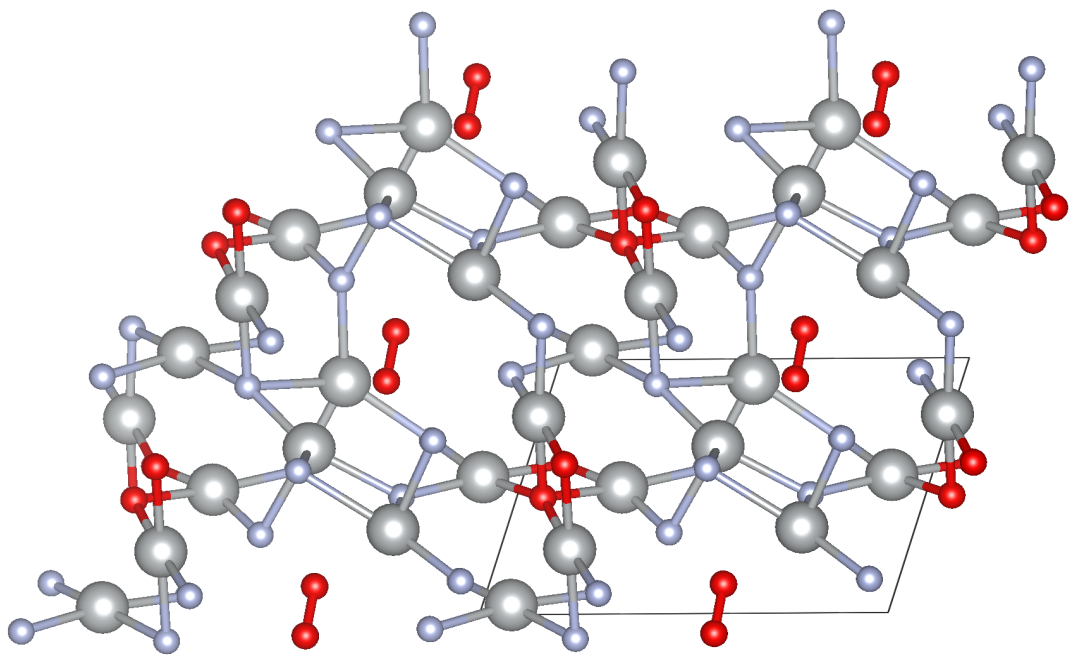

Figure 5. Crystal structure of polymorph No.62. Ag – gray, O – red, F – light blue balls.

Ag(I) sites exhibit different coordination spheres. Ag1 is ligated by 5 F atoms and 1 O atom from $O_2$ molecule in a distorted octahedral coordination; Ag4 is also surrounded by such ligands, but the polyhedron is highly irregular; same is true for Ag6; finally, Ag2 is coordinated by four F atoms and 2 trans isolated O atoms. Ag(II) sites show much more regular coordination sphere in the form of square plane, as typical for such species. Ag3, Ag5, Ag7 and Ag8 are each coordinated by 2 F and 2 isolated O ligands in the cis position. Ag(II)$_2$OF$_4^{2-}$ anion is polymeric and it adops geometry which is unprecedented in chemistry of Ag(II) (Figure 6). Oxide anions are shared between Ag(II) centers (one oxide anion between four Ag atoms) and two fluoride anions at each Ag are terminal. Such quasi-M$_2$L$_5$ stoichiometry (M = metal, L = ligand) has not yet been seen in Group 11 compounds.

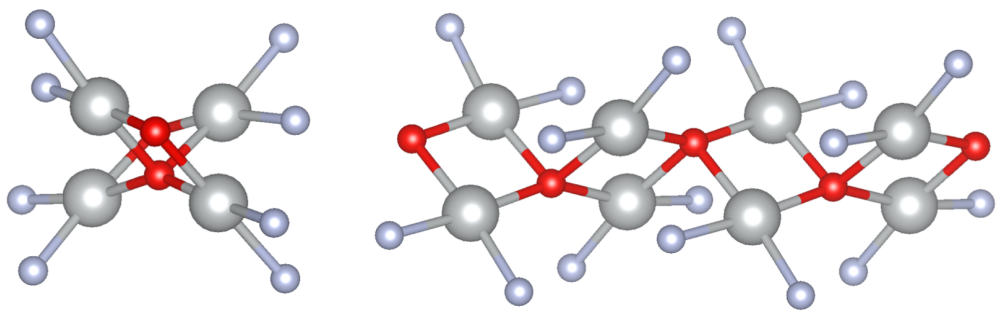

Figure 6. Two projections of the polymeric Ag(II)$_2$OF$_4^{2-}$ anion found in polymorph No.62. Left: perpendicular to the propagation direction of the chain; right – along the propagation direction of the chain. Ag – gray, O – red, F – light blue balls.

### III. Structures featuring isolated O ligands.

There are only six important structures where $O_2$ units are absent; they all fall at rather high energy with respect to our ground state structure. For example, the most stable of those (structure No.219) is found ca. 0.5 eV per formula unit above the ground state; this corresponds again to ca. 1 eV per $O_2$ molecule broken. Obviously, this value is only a fraction (about 1/5) of the O-O bond dissociation energy in an isolated molecule, since dissociation is adjoined by substantial charge transfer beteen metal and ligands and formation of stronger Ag-O bonds in the structures studied. All structures bearing isolated O ligands exhibit substantial values of magnetic moment on each independent silver site; therefore, their formulation as Ag(II)$_2$(O$^{-2}$)(F$^-$)$_2$ is appropriate.

Structure No.219 (Figure 7) contains eight independent Ag sites in the structure; each Ag is coordinated in a quasi-square planar fashion by oxide and fluoride ligands. Majority of Ag sites is coordinated by two O and 2 F ligands, some (Ag5, Ag6) in trans and some (Ag1, Ag3, Ag8) in cis configuration. One Ag site (Ag2) has three O and one F atoms in its first coordination sphere, and other two Ag sites (Ag4, Ag7) bind instead to three F and one O atoms each. The Ag-O and Ag-F bond lengths within all squares range between 2.04 Å and 2.18 Å (for O), and between 2.03 Å and 2.24 Å (F) and they are are typical of Ag(II) connections with F and O, respectively.[1,43,44,45]

Connectivity of AgL$_4$ squares in form No.219 is quite complex, as typical for compounds showing formal quasi-M$_2$L$_3$ stoichiometry and rigid [ML$_4$] coordination spheres (as exemplified by Ag$_2$O$_3$). Despite that complexity, certain similarity with Ag$_2$O$_3$ may be found. Structure No.219 consists of puckered layers of Ag atoms propagating 'parallel' to the crystallographic **ab** planes; in both forms of Ag$_2$O$_3$ such layers are flat.[46,47]

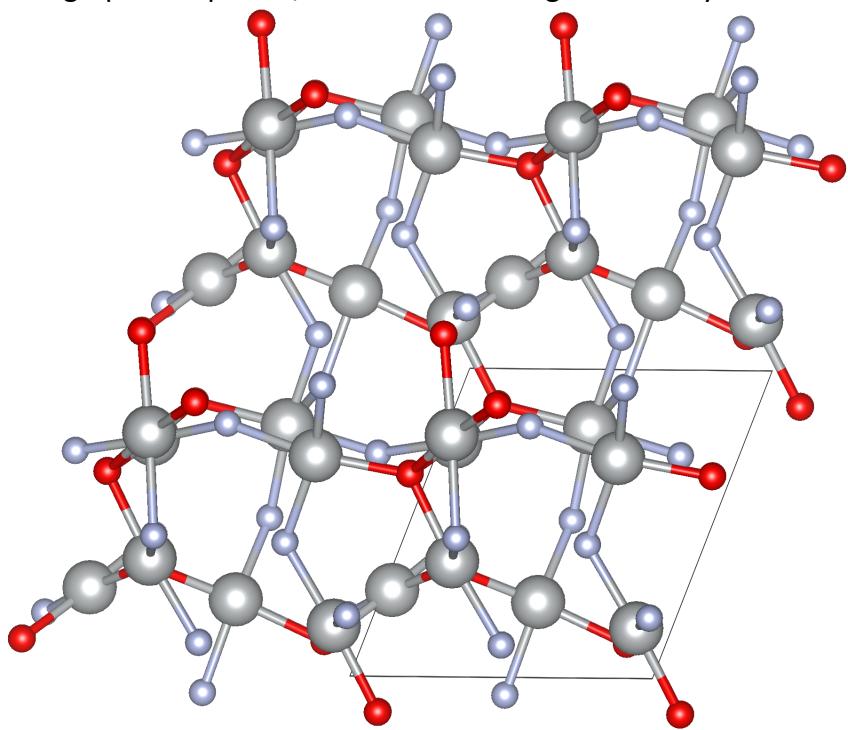

Figure 7. Crystal structure of polymorph No.219. Ag – gray, O – red, F – light blue balls.

Structure No.346 (*P*4$_1$), features ribbons in form of double chains (Figure 8). Again, each Ag is coordinated by four ligands (arranged in cis fashion); each of terminal fluoride ligands is shared between two silver atoms, while each of bridging oxide ligands is shared by four Ag atoms. Such connectivity is least energetically favoured and it also results in quite large uncompensated spin on bridging oxide anions (up to 0.4 μ$_B$). Such dramatic opening of the s$^2$p$^8$ octet must bring bad consequences for energy of this structure, which is indeed observed.

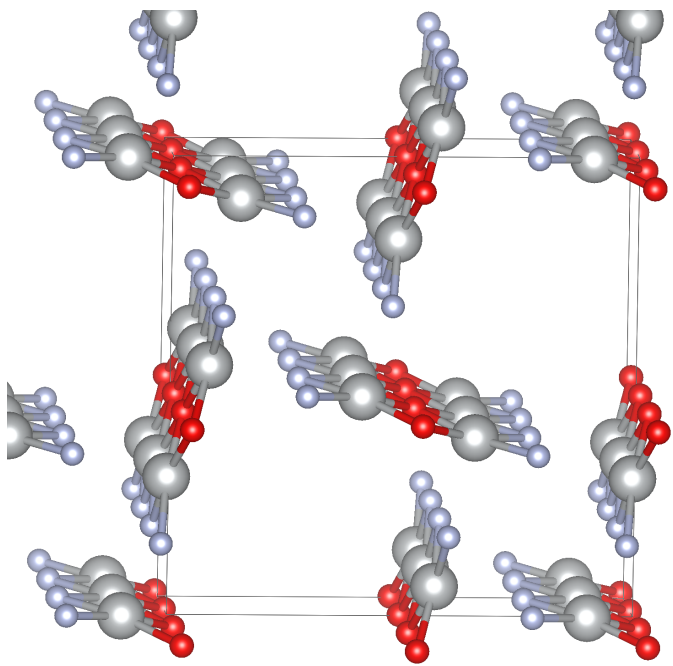

Figure 8. Crystal structure of polymorph No.346 ($P4_1$). Ag – gray, O – red, F – light blue balls.

Structure based on the $Pb_2OF_2$ polytype ($P$-4m2) (Figure 9) has preserved its high symmetry during optimization even if symmetry constrants were released. It contains silver in distorted octahedral coordination, with distances to four F atoms of 2.34–2.37 Å and two to O atoms of 2.14 and 2.37 Å. One central O atom coordinated by four Ag centers attained rather large magnetic moment of 0.78 $\mu_B$, while those at neighbouring Ag centers fallen to 0.23-0.25 $\mu_B$ level. This suggests that valence d functions of four Ag(II) sites hybridize strongly with valence functions of $sp^3$-like $O^{2-}$ anion and holes in Ag d set are partially transferred to oxygen. Other Ag centers are found in more regular coordination, with four F atoms at 2.23–2.27 Å and two O atoms at 1.96–2.08 Å; they carry negligible spin and as such they correspond to Ag(I) sites. Spin is partly trasnefrred to O atoms, and in consequence there is an uncompensated moderate magnetic moment of 1.9 $\mu_B$ in the unit cell.

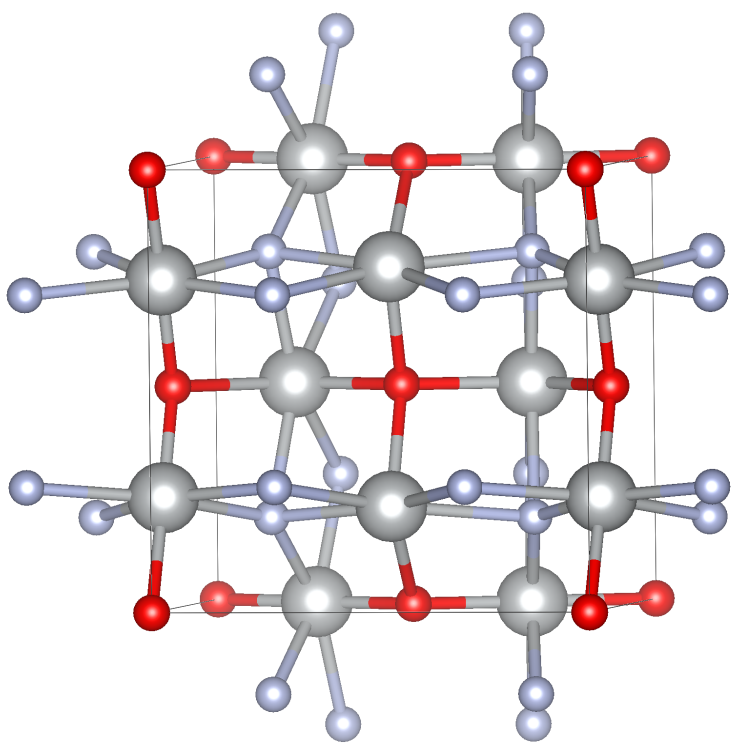

Figure 9. Crystal structure of polymorph based on the Pb$_2$OF$_2$ polytype (*P*-4m2). Ag – gray, O – red, F – light blue balls.

Structure based on the Sn$_2$OF$_2$ polytype (*C*2/m, after optimization of the original *P*2$_1$/c polytype and symmetry recognition) (Figure 10) contains two types of Ag(II) cationic sites, one with magnetic moments of 0.55 μ$_B$ (in square planar coordination with Ag–F and Ag–O bond lengths of 2.10 Å) and second with (uncommon in Ag(II) chemistry) distorted trigonal bipyramidal coordination (Ag–F 2.35–2.39 Å, Ag–O 2.08–2.10 Å) and smaller magnetic moment (0.48 μ$_B$). In both polyhedra O atoms coordinate metal in the cis fashion. This structure is ferromagnetic with substantial spin of 7.6 μ$_B$ in the cell.

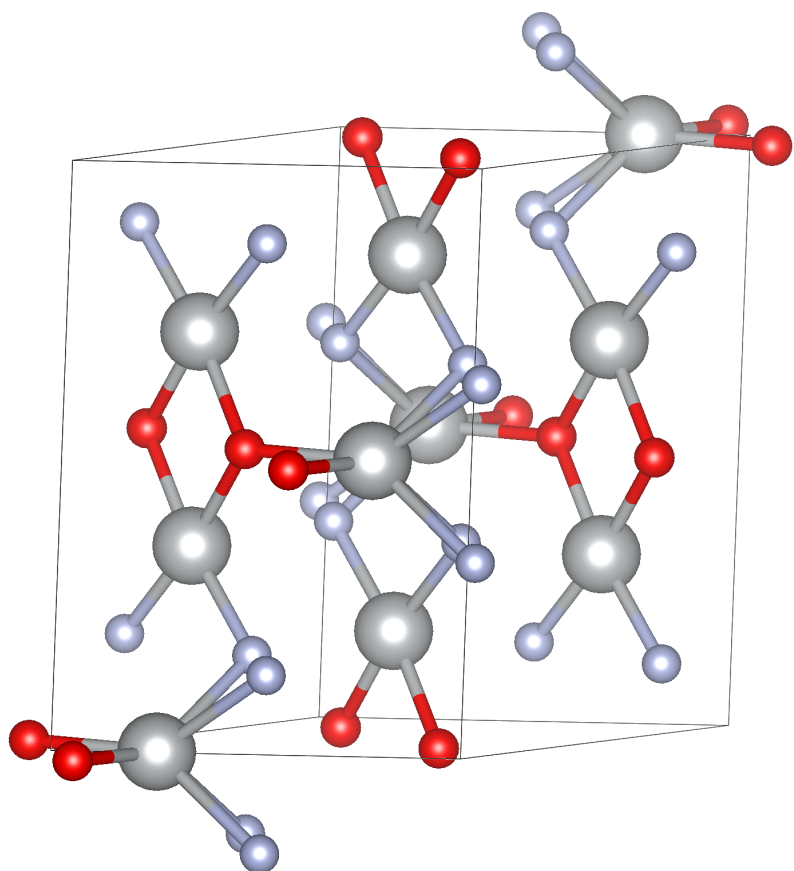

Figure 10. Crystal structure of polymorph based on the $Sn_2OF_2$ polytype ($C2/m$). Ag – gray, O – red, F – light blue balls.

The lowest-energy structure was derived from cubic $Ag_2O$ by adding F atoms to interstitial sites (Figure 11) i.e. its parent structure was a distorted fluorite type $Ag(F,O)_2$ with ¼ anionic vacancies. Optimized without any symmetry restrictions, this polytype lowers its symmetry down to $P1$. It features four types of independent Ag sites: two Ag atoms (Ag2 and Ag4) carry substantial spin (+0.26 $\mu_B$, –0.31 $\mu_B$) and they are coordinated in a square planar way by two F atoms (2.05–2.15 Å) and two O atoms in trans fashion (2.00–2.02 Å). Obviously, these are Ag(II) sites. Another Ag site (Ag1) also adopts square planar coordination but with shorter bonds (Ag–F 2.08 Å, Ag–O 1.96 Å in trans fashion) and it carries no spin. This is a low-spin Ag(III) site. Finally, the last Ag site (Ag3) has a linear coordination with two Ag–F bonds at 2.19–2.20 Å; it corresponds to a typical Ag(I). Thus, this polytype is a long sought "solid solution" of comproportionated $AgF_2$ and disproportionated AgO, preserving all essential features of the two component stoichiometries simultaneously.

Overall arrangment of spins on Ag(II) centers is antiferromagnetic, with null magnetization on the unit cell. $Ag(II)F_2O^{2-}$ sublattice is present in the form of 1D kinked chains $[Ag(II)F_2O_{2/2}{}^{2-}]$ with superexchange transmitted via O atoms. The Ag–O–Ag angle is close to 135°, which results in antiferromagnetic superexchange. The calculated magnetic superexchange constant $J_{1D}$ is –106.9 meV. This value may be compared to that of ca. –100 meV measured and computed for LT-form of $KAgF_3$ (with a similar 1D chain, albeit with F rather than O bridge, and the more open Ag–F–Ag angle of ca. 153°).[48]

It is expected that simultaneous presence of Ag(I), Ag(II) and Ag(III), together with two types of ligands (F−, O$^{2-}$) should result in rather narrow band gap at the Fermi level of this form, with an increased prospect for high-pressure metallization. We analyse density of states for this form in one of the forthcoming sections.

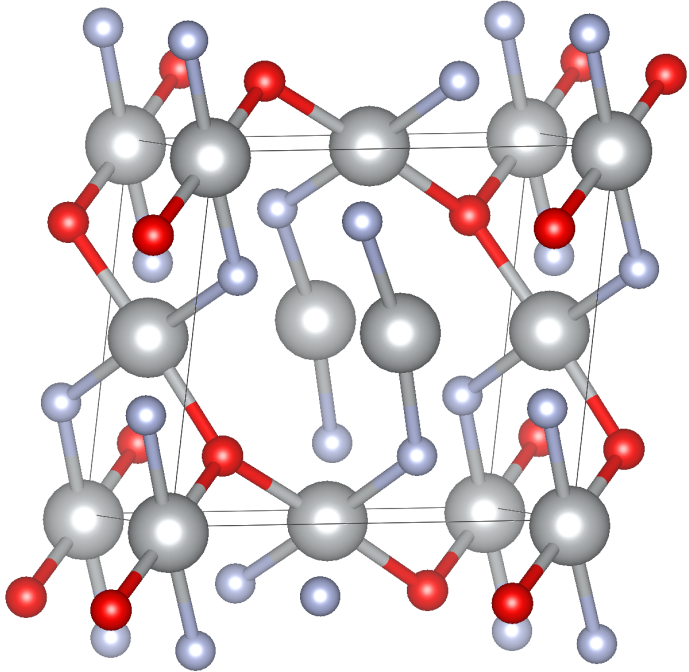

Figure 11. Crystal structure of polymorph based on the cubic Ag$_2$O by adding F atoms (*P*1). Ag – gray, O – red, F – light blue balls.

One more unique crystal structure resulted from full optimization of the second (i.e. *P*4$_2$nmc) Pb$_2$OF$_2$–type model: the *P*nnm type (Figure 12). This model is structurally reminiscent of that No.346 discussed above. It contains double chains where each O atom bridges four Ag centers. Most interestingly, one of these chains features longer Ag–F and Ag–O bonds than the other chain; the former carries unpaired electron density, the other does not. This peculiar structure turns out to the the most stable solution, since all attempts to either introduce spin on the second chain, or to eliminate spin completely, always led to higher energy.

Within the magnetic double chain, spin on O and F atoms is null, while Ag centers have substantial spin (±0.35 μ$_B$). Spin arrangement within double chain is such that Ag atoms brigded with each other via two O atoms have identical spin while those neighbouring each other via one O and one F atom have opposite spins.

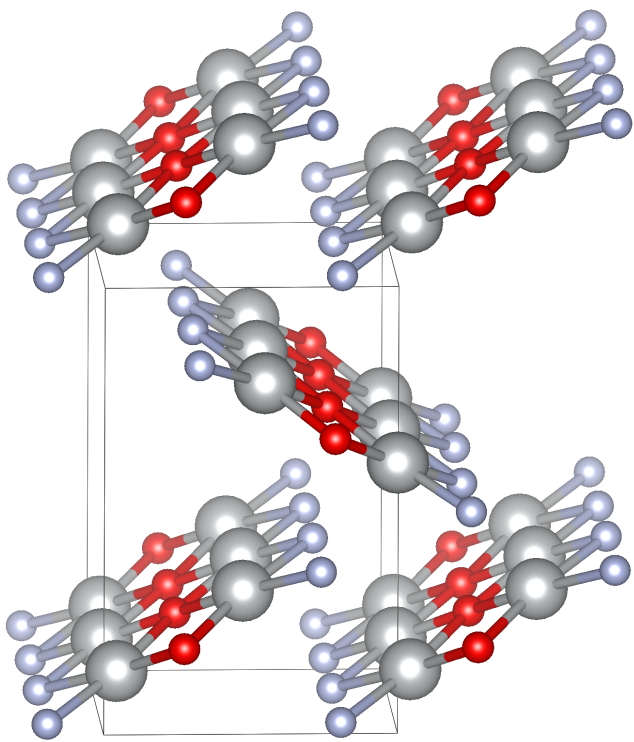

Figure 12. Crystal structure of polymorph based on the symmetry-broken Pb$_2$OF$_2$ polytype (*P*nnm). Ag – gray, O – red, F – light blue balls. One of double chains is antiferromagnetic, the other is not.

Structure No. 423 (*P*4/nmm), which has one of the most positive relative energies among all discussed here (+1.44 eV with respect to the ground state structure), has a mixed valence Ag(I)/Ag(III) character (Figure 13). All Ag(I) cations are found in quasi-cubic coordination (by either eight F or by four F and four O atoms). On the other hand, low-spin Ag(III) cations adopt a typical square planar coordination, with either four F or four O atoms in their first coordination sphere. Calculated bond lengths are typical for respective cations in such ligand environments. This structure is somewhat reminiscent of the disproportionated high-pressure HP-II structure of AgO.[23] Very high relative energy of the form No.423 is indicative of substantial energy penalty for disproportionation of Ag(II) in fluoride-rich environment. Indeed, form originating from partial F addition to Ag$_2$O (only half-disproportionated) has relative energy above the ground state structure which is about half of that for No.423.

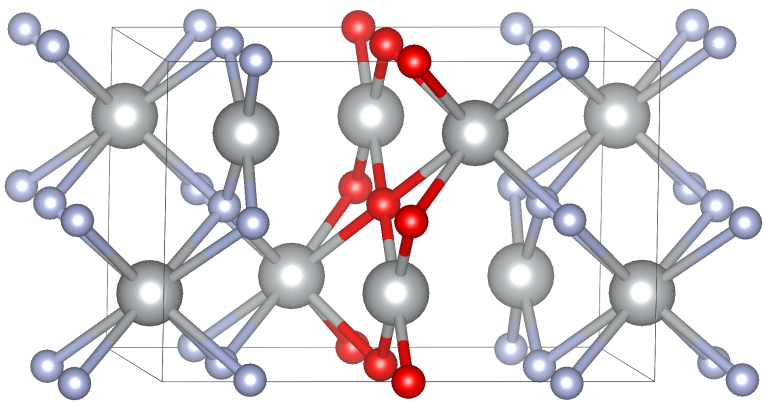

Figure 13. Crystal structure of polymorph No.423. Ag – gray, O – red, F – light blue balls.

## IV. Electronic structure of the form derived from F addition to the Ag$_2$O structure.

The previously described *P*1 polymorph (Figure 11) is characterized by the presence of silver at three different oxidation states, and of the fluoride and oxide ligands. This certainly comes with a very large span of electron energies sitting in the d bands of metal and hybridized with s,p-bands of ligands. In consequence the fundamental band gap should be much smaller than those for AgO (ca. 1 eV) or AgF$_2$ (ca. 1.6 eV). Indeed, inspection of the electronic density of states (Figure 14) shows that this form of Ag$_2$OF$_2$ is characterized by closed gap at the Fermi level when calculated with DFT+U approach. At the HSE06 level, which is known to more correctly reproduce the band gap, a band gap opens of a mere 0.245 eV between top of the valence band (predominated by O states) and bottom of the conduction band (which has substantial contribution from Ag(III) states); indeed, this value is much smaller than that calculated for AgO or AgF$_2$ using the same methodology. This also means that polymorphic form analysed here has good prospect for external-pressure induced metallization occurring via band overlap.

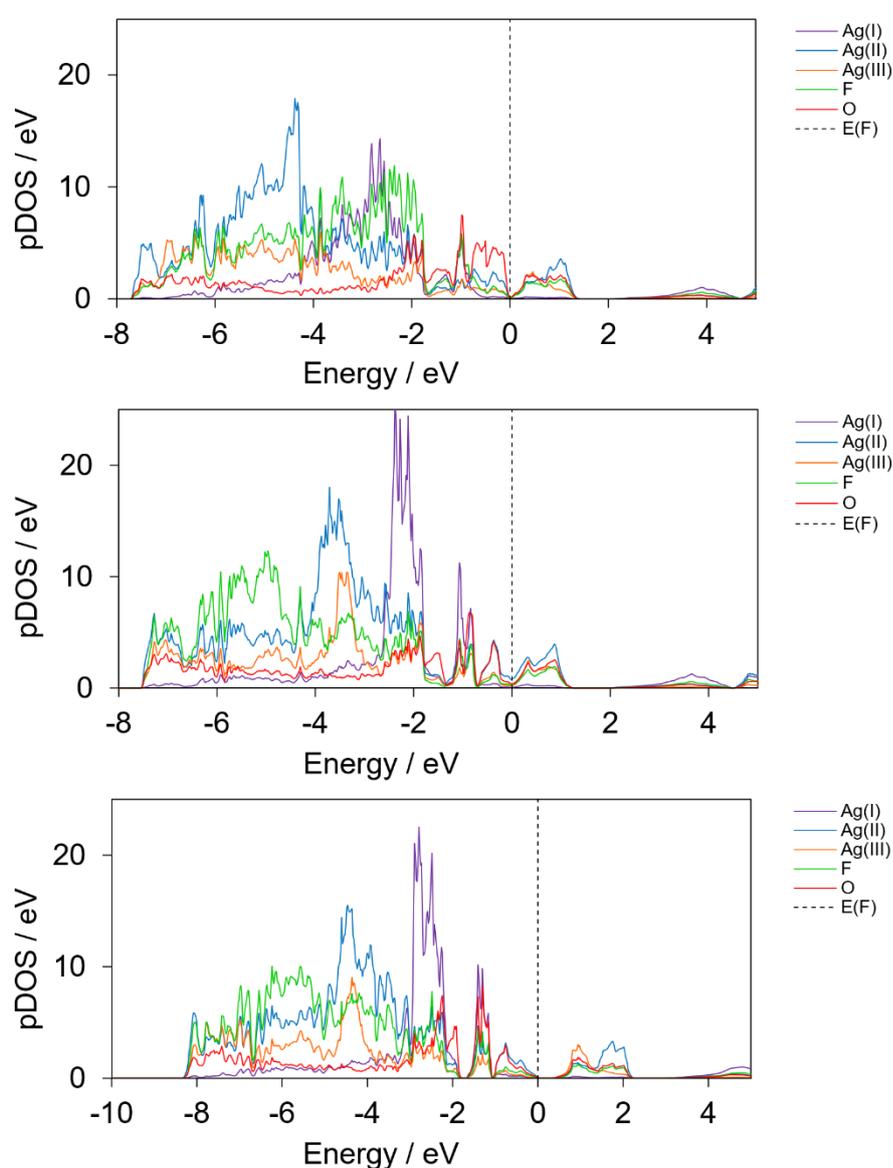

Figure 14. Electronic density of states for polymorph derived from fluorine addition to Ag$_2$O (only spin-up channel has been showed). Top: DFT+U results. Middle: SCAN meta-GGA results, and bottom: HSE06 results; both for geometry optimized at the DFT+U level.

### V. Stability of predicted structures with respect to various sets of reagents and prospect for their synthesis.

It is justified to suggest possible synthetic strategies towards $Ag_2OF_2$. Inspection of Table 1 suggests that direct synthesis from AgO and $AgF_2$ may be energetically downhill only for forms which show the presence of $O_2$ molecules in their crystal structure. Since system composed of 2 AgF and ½ $O_2$ has even lower energy, such synthesis should result in loss of oxygen and formation of AgF. We have attempted to conduct the said synthesis in the dry (mechanochemical) way, which proved successful in the past in preparation of a large numer of complex materials. Due to highly corrosive nature of $AgF_2$, we have used milling vessel and milling cylinder manufactured from pure Ni, to minimize reactivity. Unfortunately, chemical reactions and mechanical abrasion which keeps exposing fresh surface of nickel led to formation of $NiF_2$ and complex $AgNiF_3$. Indeed, in such conditions $O_2$ was lost from the sample. Use of LN2-chilled vessel did not improve the situation.

It seems that use of 1/3 $O_3$ to perform ozonolysis of 2 AgF is also a bad solution. AgF has large lattice energy and is not very reactive kinetically; even high reactivity of ozone cannot enforce formation of the desired product. Indeed, we have verified that reaction of AgF and $O_3$ (even in 100% dry gaseous or liquiefied form) does not proceed.

From the viewpoint of synthesis, $Ag_2O$ (with vacancies in its structure) and $F_2$ seem to constitute optimum set of reagents. The calculated energy of $Ag_2O$ and $F_2$ with respect to AgO and $AgF_2$ is as large as +3.898 eV, and as much as +4.387 eV with respect to 2 AgF and ½ $O_2$. This represents huge excess of free energy which may facilitate product formation. Even metastable products, such as compositions containing Ag(II) as described in section III, could in principle be prepared. If synthesis can be controlled by use of sufficiently low temperature and reduced pressure of $F_2$, to avoid evolution of $O_2$ gas, one might perhaps obtain $Ag_2OF_x$ phases, possibly including stoichiometric $Ag_2OF_2$. An alternative non-equilibrium route is offered by electrochemistry; note that one form of AgO, $Ag_2O_3$ and $Ag_3O_4$, all thermodynamically unstable with respect to $O_2$ evolution and formation of $Ag_2O$ solid residue, have been obtained using such route.

**Summary**
„Silverland" - the land of compounds of divalent silver – is rich in crystal structures and physicochemical properties which they bring.[1,49] Still, mostly due to the fact that Ag(II) is a powerful oxidizer, capable of oxidizingmost common ligands,[1,50] may even simple stoichiometries have not yet been scrutinized. Here we described the first theoretical study towards a 1:1 compound formed by AgO and $AgF_2$. Calculations show that, surprisingly, 2 Ag(I)F and ½ $O_2$ constitute ground state of such system. One predicted structure contains Ag(I) and Ag(III), just like in AgO. However, there are several structures which exhibit the presence of paramagnetic Ag(II) sites linked by fluoride and oxide anions; one of them additionally contains Ag(I) and Ag(III). The species containing paramagnetic Ag(II) are predicted to show bulk magnetic ordering and substantial spin polarization of oxide, but also partly of fluoride ligands. Magnetic properties of $Ag_2OF_2$ are consequence of strong magnetic superexchange between Ag sites, which may be realized via $F^-$ anion but also via isoelectronic $O^{2-}$ one. The latter is easier to spin-polarize than the former, since oxide's $s^2p^6$ octet is electronically softer than that of fluoride anion.

Such Ag(II) systems could possibly be obtained as metastable species while starting from $Ag_2O$ and reactive $F_2$, or using electrochemical routes. After all, chemistry is the art of manufacturing and transforming species, which most often are metastable with respect to thermodynamic sinks. Time will show whether $Ag_2OF_2$ could indeed be prepared.

**Acknowledgements**

WG acknowledges Polish National Science Center (NCN) for Beethoven project (2016/23/G/ST5/04320). This research was carried out with the support of the Interdisciplinary Centre for Mathematical and Computational Modelling (ICM), University of Warsaw under grant no. GA83-34.

**Appendix 1**
**A1. Energy convergence for generated crystal structures.**

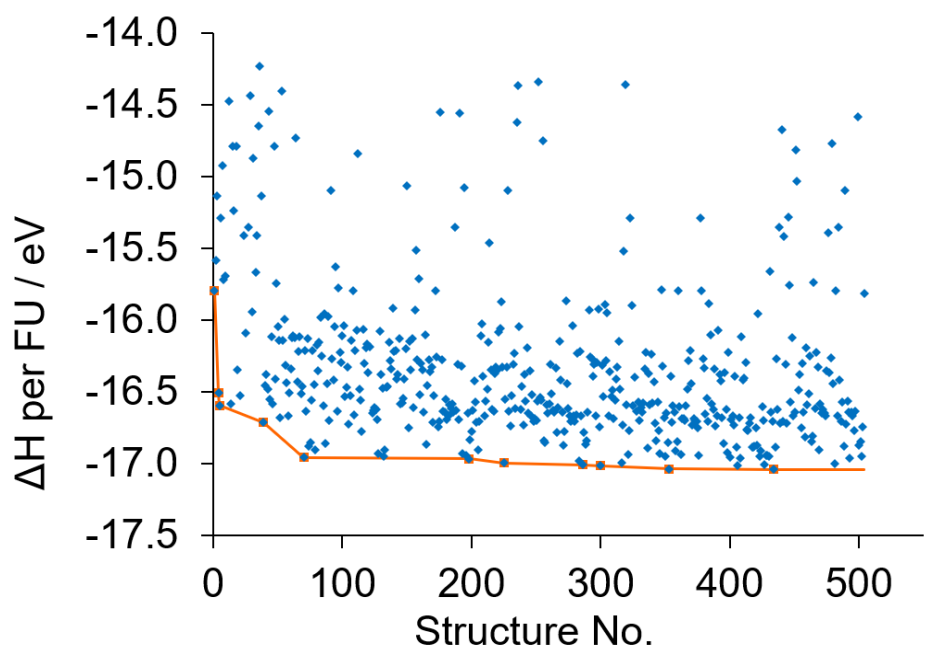

Fig. A1. Convergence of $Ag_2F_2O$ energy at 0 GPa with increasing number of structures.

**A2. Energies and volumes of key reagents and possible products.**

| System | Formula | E / eV | V | Z | E/Z | V/Z |
|---|---|---|---|---|---|---|
| $Ag_2O$ ($Cu_2O$ type) | Ag4 O2 | -18.9219 | 103.51 | 2 | -9.4609 | 51.7550 |
| $AgF_2$ (Pbca, AFM) | Ag4 F8 | -37.6107 | 161.89 | 4 | -9.4027 | 40.4725 |
| AgO (CuO type) | Ag4 O4 | -29.4474 | 103.12 | 4 | -7.3618 | 25.7800 |
| AgF (NaCl type) | Ag4 F4 | -24.1210 | 114.19 | 4 | -6.0303 | 28.5475 |
| α-$O_2$ (AFM)[51] | O4 | -20.7702 | 49.07 | 4 | -5.1926 | 12.2675 |
| α-$F_2$[52] | F8 | -15.5650 | 120.37 | 8 | -1.9456 | 15.0463 |

**A3. Crystallographic information files for the most important structures.**
It should be noticed that due to moderate value of the numer of formula units in the unit cell (up to 4 only), complexity of the stoichiometric formula $Ag_2F_2O$, and especially complex structures of the clathrate/$O_2$ polymorphs, most structures found by XtalOpt adopt *P*1 symmetry; symmetry search did not result in higher symmetry in such cases.

```
1-00007x00060-CONTCAR.cif
_symmetry_space_group_name_H-M      'P-1'
_symmetry_Int_Tables_number         2
_symmetry_cell_setting              triclinic
loop_
_symmetry_equiv_pos_as_xyz
  x,y,z
  -x,-y,-z
_cell_length_a                      4.7543
_cell_length_b                      7.7424
_cell_length_c                      8.3054
_cell_angle_alpha                   73.5695
_cell_angle_beta                    82.0311
_cell_angle_gamma                   87.6594
loop_
_atom_site_label
_atom_site_type_symbol
_atom_site_fract_x
_atom_site_fract_y
_atom_site_fract_z
_atom_site_occupancy
Ag1    Ag    0.28438    0.89149    0.62269      1.00
Ag2    Ag    0.20061   -0.02872    0.14435      1.00
Ag3    Ag    0.89935    0.58401    0.66287      1.00
Ag4    Ag    0.43189    0.27493    0.62583      1.00
F1     F     0.39808    0.57256    0.66421      1.00
F2     F     0.70033   -0.02165    0.14591      1.00
F3     F     0.21941    0.11104    0.36187      1.00
F5     F     0.93030    0.29162    0.63157      1.00
O1     O     0.80167    0.44577   -0.04819      1.00
O3     O     1.01890    0.66005   -0.02006      1.00

2-00006x00065-CONTCAR.cif
_symmetry_space_group_name_H-M      'P1'
_symmetry_Int_Tables_number         1
_symmetry_cell_setting              triclinic
loop_
_symmetry_equiv_pos_as_xyz
  x,y,z
_cell_length_a                      4.7141
_cell_length_b                      7.1668
_cell_length_c                      8.5437
_cell_angle_alpha                   101.0201
_cell_angle_beta                    96.6978
_cell_angle_gamma                   102.9560
loop_
_atom_site_label
_atom_site_type_symbol
_atom_site_fract_x
_atom_site_fract_y
_atom_site_fract_z
_atom_site_occupancy
Ag1    Ag    0.03576    0.13766    0.30011      1.00
Ag2    Ag    0.11221    0.08179    0.74625      1.00
```

```
Ag3    Ag     0.47820    0.42509    0.17457    1.00
Ag4    Ag     0.12378    0.76678    0.16399    1.00
Ag5    Ag     0.03870    0.56544    0.44720    1.00
Ag6    Ag     0.58558    0.33050    0.59367    1.00
Ag7    Ag     0.62132   -0.07776    0.46899    1.00
Ag8    Ag     0.62670    0.05216    0.02158    1.00
F1     F     -0.03208    0.41448    0.15971    1.00
F2     F      0.62658    0.10896    0.75916    1.00
F3     F      0.12064   -0.09693    0.47753    1.00
F4     F      0.54027    0.16497    0.30963    1.00
F5     F      0.63034    0.78007    0.17182    1.00
F6     F      0.54419    0.58209    0.45491    1.00
F7     F      0.11355    0.01399    0.00595    1.00
F8     F      0.07197    0.30778    0.57419    1.00
O1     O      0.16358    0.54939    0.89518    1.00
O2     O      0.63822    0.61299    0.80663    1.00
O3     O      0.31997    0.43365   -0.09161    1.00
O4     O      0.46601    0.71847    0.79701    1.00

3-00007x00032-CONTCAR.cif
_symmetry_space_group_name_H-M      'P-1'
_symmetry_Int_Tables_number         2
_symmetry_cell_setting              triclinic
loop_
_symmetry_equiv_pos_as_xyz
  x,y,z
  -x,-y,-z
_cell_length_a                      6.0983
_cell_length_b                      6.9426
_cell_length_c                      6.9626
_cell_angle_alpha                   87.2793
_cell_angle_beta                    86.3386
_cell_angle_gamma                   65.0318
loop_
_atom_site_label
_atom_site_type_symbol
_atom_site_fract_x
_atom_site_fract_y
_atom_site_fract_z
_atom_site_occupancy
Ag1    Ag     0.50246    0.11316    0.28285    1.00
Ag4    Ag    -0.12567    0.11414    0.72659    1.00
Ag5    Ag    -0.13236    0.76942    0.99395    1.00
F1     F     -0.17951    0.15886    0.11228    1.00
F2     F      0.17402    0.49867    0.19482    1.00
F3     F      0.18270    0.12710    0.49984    1.00
F7     F      0.48226    0.18614    0.85712    1.00
O1     O      0.32308    0.54025    0.60785    1.00
O3     O      0.36884    0.68425    0.52404    1.00
Ag3    Ag     0.00000    0.50000    0.50000    1.00
Ag8    Ag     0.50000    0.50000    1.00000    1.00

4-00006x00008-CONTCAR.cif
_symmetry_space_group_name_H-M      'P1'
```

```
_symmetry_Int_Tables_number        1
_symmetry_cell_setting             triclinic
loop_
_symmetry_equiv_pos_as_xyz
  x,y,z
_cell_length_a                     5.9803
_cell_length_b                     6.9604
_cell_length_c                     7.6942
_cell_angle_alpha                  98.9174
_cell_angle_beta                   96.2197
_cell_angle_gamma                  110.0817
loop_
_atom_site_label
_atom_site_type_symbol
_atom_site_fract_x
_atom_site_fract_y
_atom_site_fract_z
_atom_site_occupancy
Ag1    Ag    -0.03702    0.59186    0.40444    1.00
Ag2    Ag     0.62985    0.27778    0.59886    1.00
Ag3    Ag     0.61266    0.82115    0.56568    1.00
Ag4    Ag     0.84930    0.20391    0.02801    1.00
Ag5    Ag     0.82004    0.66496   -0.00396    1.00
Ag6    Ag     0.39638    0.88968    0.21969    1.00
Ag7    Ag     0.39090    0.35675    0.15751    1.00
Ag8    Ag     0.17924    0.00047    0.76982    1.00
F1     F      0.84207    0.64687    0.68717    1.00
F2     F      0.88979    0.13464    0.72649    1.00
F3     F      0.66693    0.70005    0.28154    1.00
F4     F      0.28358   -0.06869    0.49334    1.00
F5     F      0.53633    0.31637    0.89401    1.00
F6     F      0.11696    0.00667    0.06362    1.00
F7     F      0.69002    0.23558    0.29509    1.00
F8     F      0.11260    0.53991    0.13125    1.00
O1     O      0.30489    0.34507    0.44209    1.00
O2     O      0.56574    0.87365   -0.03169    1.00
O3     O      0.42117    0.82194    0.81685    1.00
O4     O      0.27396    0.50637    0.52155    1.00

5-00005x00089-CONTCAR.cif
_symmetry_space_group_name_H-M     'P-1'
_symmetry_Int_Tables_number        2
_symmetry_cell_setting             triclinic
loop_
_symmetry_equiv_pos_as_xyz
  x,y,z
  -x,-y,-z
_cell_length_a                     6.3169
_cell_length_b                     6.8393
_cell_length_c                     6.9250
_cell_angle_alpha                  90.9691
_cell_angle_beta                   93.0601
_cell_angle_gamma                  116.7879
loop_
```

```
_atom_site_label
_atom_site_type_symbol
_atom_site_fract_x
_atom_site_fract_y
_atom_site_fract_z
_atom_site_occupancy
Ag1    Ag    0.00424    0.22212    0.27780    1.00
Ag2    Ag    0.39348    0.15586    0.10177    1.00
Ag3    Ag    0.19837    0.25787    0.66511    1.00
Ag5    Ag    0.39606   -0.24155    0.48720    1.00
F1     F     0.40469    0.04072    0.78244    1.00
F2     F     0.56738    0.54583    0.70319    1.00
F3     F     0.18932   -0.01260    0.40358    1.00
O1     O    -0.13105    0.57797    0.06291    1.00
O2     O     0.20092    0.62527   -0.06097    1.00
F5     F     0.00000    0.00000    0.00000    1.00
F6     F     0.00000    0.50000    0.50000    1.00

6-00007x00053-CONTCAR.cif
_symmetry_space_group_name_H-M    'P1'
_symmetry_Int_Tables_number       1
_symmetry_cell_setting            triclinic
loop_
_symmetry_equiv_pos_as_xyz
  x,y,z
_cell_length_a                    6.5608
_cell_length_b                    6.5814
_cell_length_c                    6.6463
_cell_angle_alpha                 74.6354
_cell_angle_beta                  75.0780
_cell_angle_gamma                 89.7932
loop_
_atom_site_label
_atom_site_type_symbol
_atom_site_fract_x
_atom_site_fract_y
_atom_site_fract_z
_atom_site_occupancy
Ag1    Ag    0.03156    0.05368    0.44071    1.00
Ag2    Ag    0.26318    0.47552    0.36736    1.00
Ag3    Ag    0.79635    0.58594    0.46235    1.00
Ag4    Ag    0.51408   -0.06422   -0.00694    1.00
Ag5    Ag    0.71989    0.39587   -0.07707    1.00
Ag6    Ag    0.15539    0.20533   -0.03465    1.00
Ag7    Ag    0.63078    0.17811    0.39049    1.00
Ag8    Ag   -0.05387    0.74782   -0.00205    1.00
F1     F    -0.03134    0.39291    0.22035    1.00
F2     F     0.56435    0.23928    0.70566    1.00
F3     F     0.14731    0.81152    0.23094    1.00
F4     F     0.38118    0.21762    0.17610    1.00
F5     F     0.60113    0.60268    0.17964    1.00
F6     F     0.81018    0.77922    0.70080    1.00
F7     F     0.02328    0.29928    0.67421    1.00
F8     F     0.78858    0.03382    0.12464    1.00
```

| | | | | | |
|---|---|---|---|---|---|
| O1 | O | 0.27531 | 0.88784 | 0.79415 | 1.00 |
| O2 | O | 0.34729 | 0.57434 | 0.79713 | 1.00 |
| O3 | O | 0.32692 | -0.06309 | 0.58891 | 1.00 |
| O4 | O | 0.40589 | 0.63131 | 0.59569 | 1.00 |

```
7-00008x00042-CONTCAR.cif
_symmetry_space_group_name_H-M     'P1'
_symmetry_Int_Tables_number        1
_symmetry_cell_setting             triclinic
loop_
_symmetry_equiv_pos_as_xyz
  x,y,z
_cell_length_a                     5.8274
_cell_length_b                     6.2003
_cell_length_c                     8.6936
_cell_angle_alpha                  104.3269
_cell_angle_beta                   98.5853
_cell_angle_gamma                  104.9245
loop_
_atom_site_label
_atom_site_type_symbol
_atom_site_fract_x
_atom_site_fract_y
_atom_site_fract_z
_atom_site_occupancy
Ag1    Ag    0.65319    0.77763    0.47335    1.00
Ag2    Ag    0.08985    0.19575    0.77838    1.00
Ag3    Ag    0.61060    0.55158    0.07151    1.00
Ag4    Ag    0.05025    0.54558    0.59589    1.00
Ag5    Ag    0.08384    0.76834    0.00369    1.00
Ag6    Ag    0.58262    0.12660    0.26747    1.00
Ag7    Ag    0.59563   -0.05178    0.87771    1.00
Ag8    Ag    0.08536    0.35693    0.16228    1.00
F1     F     0.36597    0.72986    0.22897    1.00
F2     F     0.83308   -0.03106    0.12151    1.00
F3     F     0.82834    0.16239    0.52280    1.00
F4     F     0.84216    0.34259   -0.07285    1.00
F5     F     0.32346    0.58646    0.85266    1.00
F6     F     0.87969    0.80159    0.73867    1.00
F7     F     0.34245    0.15436    0.01550    1.00
F8     F     0.82834    0.52033    0.32736    1.00
O1     O     0.36499    0.89175    0.62750    1.00
O2     O     0.30168    0.28213    0.36234    1.00
O3     O     0.30449    0.06918    0.60918    1.00
O4     O     0.37488    0.45868    0.49150    1.00

8-00005x00019-CONTCAR.cif
_symmetry_space_group_name_H-M     'P1'
_symmetry_Int_Tables_number        1
_symmetry_cell_setting             triclinic
loop_
_symmetry_equiv_pos_as_xyz
  x,y,z
_cell_length_a                     4.6894
```

```
_cell_length_b                          7.2821
_cell_length_c                          8.3234
_cell_angle_alpha                       97.6842
_cell_angle_beta                        98.3123
_cell_angle_gamma                       98.1199
loop_
_atom_site_label
_atom_site_type_symbol
_atom_site_fract_x
_atom_site_fract_y
_atom_site_fract_z
_atom_site_occupancy
Ag1    Ag    0.10451    0.15461    0.30286     1.00
Ag2    Ag    0.14394    0.02086    0.73037     1.00
Ag3    Ag    0.49478    0.45529    0.18738     1.00
Ag4    Ag    0.06635    0.77662    0.14123     1.00
Ag5    Ag    0.09758    0.55909    0.47700     1.00
Ag6    Ag    0.69299    0.33696    0.62280     1.00
Ag7    Ag    0.65923   -0.08128    0.45643     1.00
Ag8    Ag    0.65486    0.07663    0.03656     1.00
F1     F    -0.00392    0.45931    0.19661     1.00
F2     F     0.66135    0.07246    0.75415     1.00
F3     F     0.15011    0.87691    0.44745     1.00
F4     F     0.61407    0.19474    0.32750     1.00
F5     F     0.57501    0.78819    0.15681     1.00
F6     F     0.60228    0.59016    0.47578     1.00
F7     F     0.15042    0.05317    0.01625     1.00
F8     F     0.17461    0.27466    0.58175     1.00
O1     O     0.03156    0.49708    0.86901     1.00
O2     O     0.47930    0.62901    0.80367     1.00
O3     O     0.27375    0.44924   -0.09249     1.00
O4     O     0.25608    0.68787    0.74536     1.00

9-00006x00026-CONTCAR.cif
_symmetry_space_group_name_H-M          'P1'
_symmetry_Int_Tables_number             1
_symmetry_cell_setting                  triclinic
loop_
_symmetry_equiv_pos_as_xyz
  x,y,z
_cell_length_a                          5.9209
_cell_length_b                          7.0120
_cell_length_c                          7.7932
_cell_angle_alpha                       100.8456
_cell_angle_beta                        95.8706
_cell_angle_gamma                       110.1711
loop_
_atom_site_label
_atom_site_type_symbol
_atom_site_fract_x
_atom_site_fract_y
_atom_site_fract_z
_atom_site_occupancy
Ag1    Ag    0.45442    0.57711    0.46434     1.00
```

| | | | | | |
|---|---|---|---|---|---|
| Ag2 | Ag | 0.45960 | 0.07859 | 0.53299 | 1.00 |
| Ag3 | Ag | 0.05444 | 0.77973 | 0.71437 | 1.00 |
| Ag4 | Ag | 0.87871 | 0.83341 | 0.28181 | 1.00 |
| Ag5 | Ag | 0.25883 | 0.15542 | 0.08502 | 1.00 |
| Ag6 | Ag | 0.03021 | 0.26002 | 0.65782 | 1.00 |
| Ag7 | Ag | 0.61620 | 0.51433 | 0.88613 | 1.00 |
| Ag8 | Ag | 0.89535 | 0.38854 | 0.33226 | 1.00 |
| F1 | F | 0.15300 | 0.20301 | 0.37392 | 1.00 |
| F2 | F | 0.33136 | 0.14984 | 0.79667 | 1.00 |
| F3 | F | 0.16641 | 0.74363 | 0.41782 | 1.00 |
| F4 | F | 0.59059 | 0.47246 | 0.18909 | 1.00 |
| F5 | F | 0.31430 | 0.58840 | 0.73967 | 1.00 |
| F6 | F | -0.03867 | 0.82326 | -0.00075 | 1.00 |
| F7 | F | 0.76908 | -0.07877 | 0.55758 | 1.00 |
| F8 | F | 0.72778 | 0.42904 | 0.58934 | 1.00 |
| O1 | O | 0.55357 | 0.87676 | 0.16601 | 1.00 |
| O2 | O | 0.02935 | 0.36443 | 0.05744 | 1.00 |
| O3 | O | 0.87669 | 0.32335 | -0.08401 | 1.00 |
| O4 | O | 0.53388 | 0.04759 | 0.23957 | 1.00 |

```
10-00005x00086-CONTCAR.cif
_symmetry_space_group_name_H-M      'P1'
_symmetry_Int_Tables_number         1
_symmetry_cell_setting              triclinic
loop_
_symmetry_equiv_pos_as_xyz
  x,y,z
_cell_length_a                      6.0309
_cell_length_b                      6.6812
_cell_length_c                      7.3418
_cell_angle_alpha                   94.5064
_cell_angle_beta                    93.4991
_cell_angle_gamma                   109.6050
loop_
_atom_site_label
_atom_site_type_symbol
_atom_site_fract_x
_atom_site_fract_y
_atom_site_fract_z
_atom_site_occupancy
```

| | | | | | |
|---|---|---|---|---|---|
| Ag1 | Ag | 0.81819 | -0.00990 | 0.52320 | 1.00 |
| Ag2 | Ag | 0.17593 | 0.54057 | 0.81934 | 1.00 |
| Ag3 | Ag | 0.79857 | 0.29699 | 0.09747 | 1.00 |
| Ag4 | Ag | 0.24518 | -0.06929 | 0.31453 | 1.00 |
| Ag5 | Ag | 0.62930 | 0.42653 | 0.61847 | 1.00 |
| Ag6 | Ag | 0.38195 | 0.42122 | 0.26633 | 1.00 |
| Ag7 | Ag | 0.81049 | 0.73374 | 0.05264 | 1.00 |
| Ag8 | Ag | 0.18615 | 0.09219 | 0.82872 | 1.00 |
| F1 | F | 0.07920 | 0.10466 | 0.12866 | 1.00 |
| F2 | F | 0.48710 | 0.41508 | -0.06003 | 1.00 |
| F3 | F | 0.75740 | 0.67918 | 0.34886 | 1.00 |
| F4 | F | -0.08911 | 0.74365 | 0.75776 | 1.00 |
| F5 | F | 0.85518 | 0.24246 | 0.77872 | 1.00 |
| F6 | F | 0.63280 | 0.19460 | 0.36062 | 1.00 |
| F7 | F | 0.14183 | 0.60625 | 0.13593 | 1.00 |
| F8 | F | 0.21499 | 0.19193 | 0.53248 | 1.00 |
| O1 | O | 0.42363 | 0.87950 | 0.88128 | 1.00 |
| O2 | O | 0.55656 | -0.06924 | 0.03116 | 1.00 |
| O3 | O | 0.41861 | 0.82204 | 0.53722 | 1.00 |
| O4 | O | 0.33697 | 0.61880 | 0.54663 | 1.00 |

12-00004x00067-CONTCAR.cif
```
_symmetry_space_group_name_H-M    'P1'
_symmetry_Int_Tables_number       1
_symmetry_cell_setting            triclinic
loop_
_symmetry_equiv_pos_as_xyz
  x,y,z
_cell_length_a                    6.5966
_cell_length_b                    6.6208
_cell_length_c                    6.8271
_cell_angle_alpha                 83.8341
_cell_angle_beta                  62.2202
_cell_angle_gamma                 83.2342
loop_
_atom_site_label
_atom_site_type_symbol
_atom_site_fract_x
_atom_site_fract_y
_atom_site_fract_z
_atom_site_occupancy
Ag1   Ag    0.65377    0.60746    0.03853    1.00
Ag2   Ag    0.64733   -0.01581    0.46142    1.00
Ag3   Ag    0.33882    0.17363    0.32041    1.00
Ag4   Ag    0.32849    0.72220    0.85550    1.00
Ag5   Ag    0.14965    0.79282    0.39924    1.00
Ag6   Ag    0.84384    0.50400    0.35240    1.00
Ag7   Ag    0.82755    0.00070    0.85021    1.00
Ag8   Ag    0.10867    0.17576    0.04296    1.00
F1    F    -0.04201    0.83717    0.13247    1.00
F2    F     0.24119    0.49202    0.19086    1.00
F3    F     0.43705   -0.04602    0.02355    1.00
F4    F     0.22166    0.01303    0.68860    1.00
F5    F     0.72145    0.26358    0.12443    1.00
F6    F     0.50813    0.71420    0.39390    1.00
F7    F     0.77468    0.69954    0.66773    1.00
F8    F    -0.06197    0.14787    0.46048    1.00
O1    O     0.13111    0.48072    0.80855    1.00
O2    O     0.21161    0.44781    0.60828    1.00
O3    O     0.44894    0.35877    0.78888    1.00
O4    O     0.51001    0.33286    0.58930    1.00
```

16-00002x00056-CONTCAR.cif
```
_symmetry_space_group_name_H-M    'P1'
_symmetry_Int_Tables_number       1
_symmetry_cell_setting            triclinic
loop_
_symmetry_equiv_pos_as_xyz
```

```
  x,y,z
_cell_length_a                          5.5484
_cell_length_b                          6.6689
_cell_length_c                          7.8970
_cell_angle_alpha                       99.4932
_cell_angle_beta                        105.6688
_cell_angle_gamma                       98.0589
loop_
_atom_site_label
_atom_site_type_symbol
_atom_site_fract_x
_atom_site_fract_y
_atom_site_fract_z
_atom_site_occupancy
Ag1    Ag    0.76099    0.46827   -0.09760    1.00
Ag2    Ag    0.33919    0.51877    0.09621    1.00
Ag3    Ag    0.36349   -0.08888   -0.03827    1.00
Ag4    Ag   -0.03897    0.05090    0.08406    1.00
Ag5    Ag    0.12270    0.32222    0.61074    1.00
Ag6    Ag   -0.05610    0.57775    0.28172    1.00
Ag7    Ag    0.61110    0.72389    0.58408    1.00
Ag8    Ag    0.49731    0.13129    0.32795    1.00
F1     F     0.61917    0.24928    0.08670    1.00
F2     F    -0.05099    0.77408    0.85309    1.00
F3     F     0.04962    0.25619    0.88429    1.00
F4     F     0.34575    0.80215    0.32927    1.00
F5     F     0.69642    0.77054    0.12304    1.00
F6     F     0.77511    0.43912    0.49814    1.00
F7     F     0.39249    0.59500    0.76145    1.00
F8     F     0.20103    0.33898    0.31257    1.00
O1     O     0.39761   -0.01677    0.68634    1.00
O2     O    -0.09385    0.04054    0.44606    1.00
O3     O    -0.05842    0.87930    0.50856    1.00
O4     O     0.45107    0.16738    0.66161    1.00

62-00005x00069-CONTCAR.cif
_symmetry_space_group_name_H-M         'P1'
_symmetry_Int_Tables_number             1
_symmetry_cell_setting                  triclinic
loop_
_symmetry_equiv_pos_as_xyz
  x,y,z
_cell_length_a                          5.6866
_cell_length_b                          5.9532
_cell_length_c                          9.1761
_cell_angle_alpha                       104.8310
_cell_angle_beta                        103.6734
_cell_angle_gamma                       99.5067
loop_
_atom_site_label
_atom_site_type_symbol
_atom_site_fract_x
_atom_site_fract_y
_atom_site_fract_z
```

```
_atom_site_occupancy
Ag1    Ag     0.11218   -0.07797    0.51303      1.00
Ag2    Ag     0.18046    0.03780   -0.07021      1.00
Ag3    Ag     0.68889    0.77891    0.01088      1.00
Ag4    Ag     0.52474    0.65849    0.54799      1.00
Ag5    Ag     0.65229    0.25655    0.83109      1.00
Ag6    Ag     0.76448    0.33787    0.27967      1.00
Ag7    Ag     0.18639    0.54137    0.09891      1.00
Ag8    Ag     0.14102    0.49787    0.75088      1.00
F1     F      0.41379   -0.09581    0.74685      1.00
F2     F      0.89839    0.13343    0.07388      1.00
F3     F     -0.01695    0.68641    0.25125      1.00
F4     F      0.32960    0.32018    0.60006      1.00
F5     F      0.44813   -0.05479    0.11055      1.00
F6     F     -0.08378    0.57105    0.55848      1.00
F7     F      0.89513    0.08156    0.72707      1.00
F8     F      0.40837    0.46680    0.28378      1.00
O1     O     -0.07669    0.58983   -0.08925      1.00
O2     O      0.41517    0.46034   -0.06338      1.00
O3     O      0.67713   -0.07821    0.41152      1.00
O4     O      0.61501    0.11615    0.42695      1.00

171-00003x00027-CONTCAR.cif
_symmetry_space_group_name_H-M       'P1'
_symmetry_Int_Tables_number          1
_symmetry_cell_setting               triclinic
loop_
_symmetry_equiv_pos_as_xyz
  x,y,z
_cell_length_a                       5.8278
_cell_length_b                       6.6342
_cell_length_c                       7.1522
_cell_angle_alpha                    76.3817
_cell_angle_beta                     85.6972
_cell_angle_gamma                    87.5402
loop_
_atom_site_label
_atom_site_type_symbol
_atom_site_fract_x
_atom_site_fract_y
_atom_site_fract_z
_atom_site_occupancy
Ag1    Ag     0.10520    0.59066    0.66674      1.00
Ag2    Ag     0.09147    0.31156    0.15341      1.00
Ag3    Ag     0.50652    0.89669    0.36055      1.00
Ag4    Ag    -0.04541    0.78670    0.20081      1.00
Ag5    Ag     0.57309    0.25686   -0.06470      1.00
Ag6    Ag     0.11000    0.09194    0.64021      1.00
Ag7    Ag     0.56603    0.39767    0.35471      1.00
Ag8    Ag     0.52606    0.83755    0.86736      1.00
F1     F      0.86736    0.85306    0.72535      1.00
F2     F      0.18052   -0.02779    0.01696      1.00
F3     F      0.30454    0.52538   -0.03949      1.00
F4     F      0.80271    0.48255    0.04903      1.00
```

| | | | | | |
|---|---|---|---|---|---|
| F5 | F | 0.86336 | 0.15205 | 0.36648 | 1.00 |
| F6 | F | 0.36616 | 0.30526 | 0.65756 | 1.00 |
| F7 | F | 0.21791 | 0.61305 | 0.32926 | 1.00 |
| F8 | F | 0.88470 | 0.32662 | 0.70321 | 1.00 |
| O1 | O | 0.36298 | 0.17566 | 0.30601 | 1.00 |
| O2 | O | 0.67403 | -0.06359 | 0.09845 | 1.00 |
| O3 | O | 0.72597 | 0.64678 | 0.40355 | 1.00 |
| O4 | O | 0.33795 | 0.84134 | 0.62070 | 1.00 |

```
219-00009x00010-CONTCAR.cif
_symmetry_space_group_name_H-M     'P1'
_symmetry_Int_Tables_number        1
_symmetry_cell_setting             triclinic
loop_
_symmetry_equiv_pos_as_xyz
  x,y,z
_cell_length_a                     6.4482
_cell_length_b                     6.8702
_cell_length_c                     7.5698
_cell_angle_alpha                  111.0676
_cell_angle_beta                   107.7371
_cell_angle_gamma                  93.7575
loop_
_atom_site_label
_atom_site_type_symbol
_atom_site_fract_x
_atom_site_fract_y
_atom_site_fract_z
_atom_site_occupancy
```

| | | | | | |
|---|---|---|---|---|---|
| Ag1 | Ag | -0.02880 | 0.00664 | 0.28012 | 1.00 |
| Ag2 | Ag | -0.02651 | -0.01063 | 0.78598 | 1.00 |
| Ag3 | Ag | 0.81015 | 0.55976 | 0.37552 | 1.00 |
| Ag4 | Ag | 0.52675 | 0.27869 | 0.81167 | 1.00 |
| Ag5 | Ag | -0.04631 | 0.48059 | 0.76344 | 1.00 |
| Ag6 | Ag | 0.47456 | 0.14437 | 0.24551 | 1.00 |
| Ag7 | Ag | 0.39588 | 0.70623 | 0.71524 | 1.00 |
| Ag8 | Ag | 0.11766 | 0.41199 | 0.16358 | 1.00 |
| F1 | F | 0.82649 | 0.89383 | 0.45131 | 1.00 |
| F2 | F | 0.19371 | 0.29106 | 0.72351 | 1.00 |
| F3 | F | 0.46751 | 0.11218 | 0.51158 | 1.00 |
| F4 | F | 0.38560 | 0.49384 | 0.43264 | 1.00 |
| F5 | F | 0.08767 | 0.72032 | 0.19788 | 1.00 |
| F6 | F | 0.54122 | 0.46315 | 0.10592 | 1.00 |
| F7 | F | 0.72759 | 0.68757 | 0.81724 | 1.00 |
| F8 | F | 0.46483 | 0.00412 | -0.05702 | 1.00 |
| O1 | O | 0.86210 | 0.28083 | 0.88543 | 1.00 |
| O2 | O | 0.12879 | 0.09571 | 0.11262 | 1.00 |
| O3 | O | 0.06209 | 0.69113 | 0.65879 | 1.00 |
| O4 | O | 0.81045 | 0.25415 | 0.35663 | 1.00 |

```
data_Ag2F2O--Ag2O+F2--symP1
_symmetry_space_group_name_H-M     'P1'
_symmetry_Int_Tables_number        1
_symmetry_cell_setting             triclinic
```

```
loop_
_symmetry_equiv_pos_as_xyz
  x,y,z
_cell_length_a                    5.4683
_cell_length_b                    4.5562
_cell_length_c                    5.7083
_cell_angle_alpha                 100.9360
_cell_angle_beta                  83.5130
_cell_angle_gamma                 98.5835
loop_
_atom_site_label
_atom_site_type_symbol
_atom_site_fract_x
_atom_site_fract_y
_atom_site_fract_z
_atom_site_occupancy
Ag1    Ag    0.01583    0.00761    0.00661    1.00
Ag2    Ag    0.51279    0.51044    0.00464    1.00
Ag3    Ag    0.49305    0.00111    0.48201    1.00
Ag4    Ag    0.01107    0.51104    0.50587    1.00
O1     O     0.20708    0.34948    0.20164    1.00
O2     O     0.82690    0.66116    0.81808    1.00
F1     F     0.18416    0.23719    0.65921    1.00
F2     F     0.83042    0.78482    0.35927    1.00
F3     F     0.69438    0.19488    0.13577    1.00
F4     F     0.33932    0.82727    0.86690    1.00

Ag2F2O__Pb2OF2_P42nmc_type_Pnnm.cif
_symmetry_space_group_name_H-M    'PNNM'
_symmetry_Int_Tables_number       58
_symmetry_cell_setting            orthorhombic
loop_
_symmetry_equiv_pos_as_xyz
  x,y,z
  -x,-y,z
  -x+1/2,y+1/2,-z+1/2
  x+1/2,-y+1/2,-z+1/2
  -x,-y,-z
  x,y,-z
  x+1/2,-y+1/2,z+1/2
  -x+1/2,y+1/2,z+1/2
_cell_length_a                    5.4072
_cell_length_b                    8.4362
_cell_length_c                    3.0820
_cell_angle_alpha                 90.0000
_cell_angle_beta                  90.0000
_cell_angle_gamma                 90.0000
loop_
_atom_site_label
_atom_site_type_symbol
_atom_site_fract_x
_atom_site_fract_y
_atom_site_fract_z
```

```
_atom_site_occupancy
Ag1    Ag   -0.22590   0.41046   0.50000   0.01267   1.00
O1     O     0.00000   0.50000   1.00000   0.01267   1.00
F1     F     0.53683   0.33111   1.00000   0.01267   1.00

Ag2F2O__Sn2OF2_P21c_type_C2m.cif
_symmetry_space_group_name_H-M      'C2/M'
_symmetry_Int_Tables_number         12
_symmetry_cell_setting              monoclinic
loop_
_symmetry_equiv_pos_as_xyz
  x,y,z
  -x,y,-z
  -x,-y,-z
  x,-y,z
  x+1/2,y+1/2,z
  -x+1/2,y+1/2,-z
  -x+1/2,-y+1/2,-z
  x+1/2,-y+1/2,z
_cell_length_a                      6.0149
_cell_length_b                      8.7016
_cell_length_c                      9.0487
_cell_angle_alpha                   90.0000
_cell_angle_beta                    144.2074
_cell_angle_gamma                   90.0000
loop_
_atom_site_label
_atom_site_type_symbol
_atom_site_fract_x
_atom_site_fract_y
_atom_site_fract_z
_atom_site_occupancy
F1    F    -1.17468   -0.15127   -0.74547   1.00
Ag1   Ag   -0.29203    0.00000   -0.02507   1.00
O1    O    -0.70585    0.00000   -0.74184   1.00
Ag2   Ag   -1.00000   -0.31857   -0.50000   1.00

346-00004x00041-CONTCAR.cif
_symmetry_space_group_name_H-M      'I41'
_symmetry_Int_Tables_number         80
_symmetry_cell_setting              tetragonal
loop_
_symmetry_equiv_pos_as_xyz
  x,y,z
  -x+1/2,-y+1/2,z+1/2
  -y,x+1/2,z+1/4
  y+1/2,-x,z+3/4
  x+1/2,y+1/2,z+1/2
  -x,-y,z
  -y+1/2,x,z+3/4
  y,-x+1/2,z+1/4
_cell_length_a                      11.2041
_cell_length_b                      11.2041
_cell_length_c                      3.0367
```

```
_cell_angle_alpha              90.0000
_cell_angle_beta               90.0000
_cell_angle_gamma              90.0000
loop_
_atom_site_label
_atom_site_type_symbol
_atom_site_fract_x
_atom_site_fract_y
_atom_site_fract_z
_atom_site_occupancy
Ag1    Ag    0.12163    0.04421    0.30700    1.00
F1     F     0.74849    0.56359    0.30862    1.00
O1     O     0.00000    0.00000    0.80275    1.00

Ag2F2O__Pb2OF2_P-4m2_type_P-4m2.cif
_symmetry_space_group_name_H-M     'P-4M2'
_symmetry_Int_Tables_number        115
_symmetry_cell_setting             tetragonal
loop_
_symmetry_equiv_pos_as_xyz
  x,y,z
  -x,-y,z
  y,-x,-z
  -y,x,-z
  x,-y,z
  -x,y,z
  y,x,-z
  -y,-x,-z
_cell_length_a                 7.1422
_cell_length_b                 7.1422
_cell_length_c                 5.3563
_cell_angle_alpha              90.0000
_cell_angle_beta               90.0000
_cell_angle_gamma              90.0000
loop_
_atom_site_label
_atom_site_type_symbol
_atom_site_fract_x
_atom_site_fract_y
_atom_site_fract_z
_atom_site_occupancy
Ag1    Ag    0.25504    0.00000    0.18820    1.00
Ag2    Ag    0.22825    0.50000    0.33107    1.00
O1     O     0.00000    0.00000    0.00000    1.00
O2     O     0.50000    0.50000    0.50000    1.00
O3     O     0.00000    0.50000    0.64338    1.00
F1     F     0.28193    0.28193    0.00000    1.00
F2     F     0.19930    0.19930    0.50000    1.00

423-00010x00006-CONTCAR.cif
_symmetry_space_group_name_H-M     'P4/NMM'
_symmetry_Int_Tables_number        129
_symmetry_cell_setting             tetragonal
loop_
```

```
_symmetry_equiv_pos_as_xyz
  x,y,z
  -x,-y,z
  -y+1/2,x+1/2,z
  y+1/2,-x+1/2,z
  -x+1/2,y+1/2,-z
  x+1/2,-y+1/2,-z
  y,x,-z
  -y,-x,-z
  -x+1/2,-y+1/2,-z
  x+1/2,y+1/2,-z
  y,-x,-z
  -y,x,-z
  x,-y,z
  -x,y,z
  -y+1/2,-x+1/2,z
  y+1/2,x+1/2,z
_cell_length_a                   5.5785
_cell_length_b                   5.5785
_cell_length_c                   8.6018
_cell_angle_alpha                90.0000
_cell_angle_beta                 90.0000
_cell_angle_gamma                90.0000
loop_
_atom_site_label
_atom_site_type_symbol
_atom_site_fract_x
_atom_site_fract_y
_atom_site_fract_z
_atom_site_occupancy
F1    F    0.25479  -0.24520   0.17770    1.00
Ag1   Ag   0.50000   0.50000   0.50000    1.00
Ag2   Ag   0.50000   0.50000   0.00000    1.00
Ag3   Ag   0.50000  -0.00000   0.34475    1.00
Ag4   Ag   0.50000  -0.00000   0.81040    1.00
O1    O    0.25000  -0.25000   0.50000    1.00

461-00001x00003-CONTCAR.cif
_symmetry_space_group_name_H-M    'FMMM'
_symmetry_Int_Tables_number       69
_symmetry_cell_setting            orthorhombic
loop_
_symmetry_equiv_pos_as_xyz
  x,y,z
  -x,-y,z
  -x,y,-z
  x,-y,-z
  -x,-y,-z
  x,y,-z
  x,-y,z
  -x,y,z
  x,y+1/2,z+1/2
  -x,-y+1/2,z+1/2
  -x,y+1/2,-z+1/2
```

```
  x,-y+1/2,-z+1/2
  -x,-y+1/2,-z+1/2
  x,y+1/2,-z+1/2
  x,-y+1/2,z+1/2
  -x,y+1/2,z+1/2
  x+1/2,y,z+1/2
  -x+1/2,-y,z+1/2
  -x+1/2,y,-z+1/2
  x+1/2,-y,-z+1/2
  -x+1/2,-y,-z+1/2
  x+1/2,y,-z+1/2
  x+1/2,-y,z+1/2
  -x+1/2,y,z+1/2
  x+1/2,y+1/2,z
  -x+1/2,-y+1/2,z
  -x+1/2,y+1/2,-z
  x+1/2,-y+1/2,-z
  -x+1/2,-y+1/2,-z
  x+1/2,y+1/2,-z
  x+1/2,-y+1/2,z
  -x+1/2,y+1/2,z
_cell_length_a                      6.5344
_cell_length_b                      7.4903
_cell_length_c                      5.7192
_cell_angle_alpha                   90.0000
_cell_angle_beta                    90.0000
_cell_angle_gamma                   90.0000
loop_
_atom_site_label
_atom_site_type_symbol
_atom_site_fract_x
_atom_site_fract_y
_atom_site_fract_z
_atom_site_occupancy
Ag1    Ag    0.25000    0.00000    0.25000    1.00
F1     F     0.25000    0.25000   -0.00000    1.00
O1     O     0.00000    0.00000    0.00000    1.00
```

**CIF files of reference compounds (substrates)**

```
Ag2O--Cu2O-type.cif
_audit_creation_method FINDSYM
_cell_length_a     4.6952010000
_cell_length_b     4.6952010000
_cell_length_c     4.6952010000
_cell_angle_alpha 90.0000000000
_cell_angle_beta  90.0000000000
_cell_angle_gamma 90.0000000000
_cell_volume       103.5052948881
_symmetry_space_group_name_H-M "P 42/n -3 2/m (origin choice 2)"
_symmetry_Int_Tables_number 224
_space_group.reference_setting '224:-P 4bc 2bc 3'
_space_group.transform_Pp_abc a,b,c;0,0,0
```

```
loop_
_space_group_symop_id
_space_group_symop_operation_xyz
1 x,y,z
2 x,-y+1/2,-z+1/2
3 -x+1/2,y,-z+1/2
4 -x+1/2,-y+1/2,z
5 y,z,x
6 y,-z+1/2,-x+1/2
7 -y+1/2,z,-x+1/2
8 -y+1/2,-z+1/2,x
9 z,x,y
10 z,-x+1/2,-y+1/2
11 -z+1/2,x,-y+1/2
12 -z+1/2,-x+1/2,y
13 -y,-x,-z
14 -y,x+1/2,z+1/2
15 y+1/2,-x,z+1/2
16 y+1/2,x+1/2,-z
17 -x,-z,-y
18 -x,z+1/2,y+1/2
19 x+1/2,-z,y+1/2
20 x+1/2,z+1/2,-y
21 -z,-y,-x
22 -z,y+1/2,x+1/2
23 z+1/2,-y,x+1/2
24 z+1/2,y+1/2,-x
25 -x,-y,-z
26 -x,y+1/2,z+1/2
27 x+1/2,-y,z+1/2
28 x+1/2,y+1/2,-z
29 -y,-z,-x
30 -y,z+1/2,x+1/2
31 y+1/2,-z,x+1/2
32 y+1/2,z+1/2,-x
33 -z,-x,-y
34 -z,x+1/2,y+1/2
35 z+1/2,-x,y+1/2
36 z+1/2,x+1/2,-y
37 y,x,z
38 y,-x+1/2,-z+1/2
39 -y+1/2,x,-z+1/2
40 -y+1/2,-x+1/2,z
41 x,z,y
42 x,-z+1/2,-y+1/2
43 -x+1/2,z,-y+1/2
44 -x+1/2,-z+1/2,y
45 z,y,x
46 z,-y+1/2,-x+1/2
47 -z+1/2,y,-x+1/2
48 -z+1/2,-y+1/2,x
loop_
_atom_site_label
_atom_site_type_symbol
```

```
_atom_site_symmetry_multiplicity
_atom_site_Wyckoff_label
_atom_site_fract_x
_atom_site_fract_y
_atom_site_fract_z
     _atom_site_occupancy
     _atom_site_fract_symmform
Ag1 Ag   4 b 0.00000 0.00000 0.00000 1.00000 0,0,0
O1  O    2 a 0.25000 0.25000 0.25000 1.00000 0,0,0

AgF2--Pbca-AFM.cif
_audit_creation_method FINDSYM
_cell_length_a    5.5000660000
_cell_length_b    5.8254950000
_cell_length_c    5.0527050000
_cell_angle_alpha 90.0000000000
_cell_angle_beta  90.0000000000
_cell_angle_gamma 90.0000000000
_cell_volume      161.8917351044
_symmetry_space_group_name_H-M "P 21/b 21/c 21/a"
_symmetry_Int_Tables_number 61
_space_group.reference_setting '061:-P 2ac 2ab'
_space_group.transform_Pp_abc a,b,c;0,0,0
loop_
_space_group_symop_id
_space_group_symop_operation_xyz
1 x,y,z
2 x+1/2,-y+1/2,-z
3 -x,y+1/2,-z+1/2
4 -x+1/2,-y,z+1/2
5 -x,-y,-z
6 -x+1/2,y+1/2,z
7 x,-y+1/2,z+1/2
8 x+1/2,y,-z+1/2
loop_
_atom_site_label
_atom_site_type_symbol
_atom_site_symmetry_multiplicity
_atom_site_Wyckoff_label
_atom_site_fract_x
_atom_site_fract_y
_atom_site_fract_z
     _atom_site_occupancy
     _atom_site_fract_symmform
Ag1 Ag   4 b 0.00000 0.00000 0.50000 1.00000 0,0,0
F1  F    8 c 0.30552 0.36811 0.18355 1.00000 Dx,Dy,Dz

AgF--NaCl-type.cif
_audit_creation_method FINDSYM
_cell_length_a    4.8514380000
_cell_length_b    4.8514380000
_cell_length_c    4.8514380000
_cell_angle_alpha 90.0000000000
_cell_angle_beta  90.0000000000
```

```
_cell_angle_gamma 90.0000000000
_cell_volume      114.1856311551
_symmetry_space_group_name_H-M "F 4/m -3 2/m"
_symmetry_Int_Tables_number 225
_space_group.reference_setting '225:-F 4 2 3'
_space_group.transform_Pp_abc a,b,c;0,0,0
loop_
_space_group_symop_id
_space_group_symop_operation_xyz
1 x,y,z
2 x,-y,-z
3 -x,y,-z
4 -x,-y,z
5 y,z,x
6 y,-z,-x
7 -y,z,-x
8 -y,-z,x
9 z,x,y
10 z,-x,-y
11 -z,x,-y
12 -z,-x,y
13 -y,-x,-z
14 -y,x,z
15 y,-x,z
16 y,x,-z
17 -x,-z,-y
18 -x,z,y
19 x,-z,y
20 x,z,-y
21 -z,-y,-x
22 -z,y,x
23 z,-y,x
24 z,y,-x
25 -x,-y,-z
26 -x,y,z
27 x,-y,z
28 x,y,-z
29 -y,-z,-x
30 -y,z,x
31 y,-z,x
32 y,z,-x
33 -z,-x,-y
34 -z,x,y
35 z,-x,y
36 z,x,-y
37 y,x,z
38 y,-x,-z
39 -y,x,-z
40 -y,-x,z
41 x,z,y
42 x,-z,-y
43 -x,z,-y
44 -x,-z,y
45 z,y,x
```

```
46  z,-y,-x
47  -z,y,-x
48  -z,-y,x
49  x,y+1/2,z+1/2
50  x,-y+1/2,-z+1/2
51  -x,y+1/2,-z+1/2
52  -x,-y+1/2,z+1/2
53  y,z+1/2,x+1/2
54  y,-z+1/2,-x+1/2
55  -y,z+1/2,-x+1/2
56  -y,-z+1/2,x+1/2
57  z,x+1/2,y+1/2
58  z,-x+1/2,-y+1/2
59  -z,x+1/2,-y+1/2
60  -z,-x+1/2,y+1/2
61  -y,-x+1/2,-z+1/2
62  -y,x+1/2,z+1/2
63  y,-x+1/2,z+1/2
64  y,x+1/2,-z+1/2
65  -x,-z+1/2,-y+1/2
66  -x,z+1/2,y+1/2
67  x,-z+1/2,y+1/2
68  x,z+1/2,-y+1/2
69  -z,-y+1/2,-x+1/2
70  -z,y+1/2,x+1/2
71  z,-y+1/2,x+1/2
72  z,y+1/2,-x+1/2
73  -x,-y+1/2,-z+1/2
74  -x,y+1/2,z+1/2
75  x,-y+1/2,z+1/2
76  x,y+1/2,-z+1/2
77  -y,-z+1/2,-x+1/2
78  -y,z+1/2,x+1/2
79  y,-z+1/2,x+1/2
80  y,z+1/2,-x+1/2
81  -z,-x+1/2,-y+1/2
82  -z,x+1/2,y+1/2
83  z,-x+1/2,y+1/2
84  z,x+1/2,-y+1/2
85  y,x+1/2,z+1/2
86  y,-x+1/2,-z+1/2
87  -y,x+1/2,-z+1/2
88  -y,-x+1/2,z+1/2
89  x,z+1/2,y+1/2
90  x,-z+1/2,-y+1/2
91  -x,z+1/2,-y+1/2
92  -x,-z+1/2,y+1/2
93  z,y+1/2,x+1/2
94  z,-y+1/2,-x+1/2
95  -z,y+1/2,-x+1/2
96  -z,-y+1/2,x+1/2
97  x+1/2,y,z+1/2
98  x+1/2,-y,-z+1/2
99  -x+1/2,y,-z+1/2
```

```
100 -x+1/2,-y,z+1/2
101 y+1/2,z,x+1/2
102 y+1/2,-z,-x+1/2
103 -y+1/2,z,-x+1/2
104 -y+1/2,-z,x+1/2
105 z+1/2,x,y+1/2
106 z+1/2,-x,-y+1/2
107 -z+1/2,x,-y+1/2
108 -z+1/2,-x,y+1/2
109 -y+1/2,-x,-z+1/2
110 -y+1/2,x,z+1/2
111 y+1/2,-x,z+1/2
112 y+1/2,x,-z+1/2
113 -x+1/2,-z,-y+1/2
114 -x+1/2,z,y+1/2
115 x+1/2,-z,y+1/2
116 x+1/2,z,-y+1/2
117 -z+1/2,-y,-x+1/2
118 -z+1/2,y,x+1/2
119 z+1/2,-y,x+1/2
120 z+1/2,y,-x+1/2
121 -x+1/2,-y,-z+1/2
122 -x+1/2,y,z+1/2
123 x+1/2,-y,z+1/2
124 x+1/2,y,-z+1/2
125 -y+1/2,-z,-x+1/2
126 -y+1/2,z,x+1/2
127 y+1/2,-z,x+1/2
128 y+1/2,z,-x+1/2
129 -z+1/2,-x,-y+1/2
130 -z+1/2,x,y+1/2
131 z+1/2,-x,y+1/2
132 z+1/2,x,-y+1/2
133 y+1/2,x,z+1/2
134 y+1/2,-x,-z+1/2
135 -y+1/2,x,-z+1/2
136 -y+1/2,-x,z+1/2
137 x+1/2,z,y+1/2
138 x+1/2,-z,-y+1/2
139 -x+1/2,z,-y+1/2
140 -x+1/2,-z,y+1/2
141 z+1/2,y,x+1/2
142 z+1/2,-y,-x+1/2
143 -z+1/2,y,-x+1/2
144 -z+1/2,-y,x+1/2
145 x+1/2,y+1/2,z
146 x+1/2,-y+1/2,-z
147 -x+1/2,y+1/2,-z
148 -x+1/2,-y+1/2,z
149 y+1/2,z+1/2,x
150 y+1/2,-z+1/2,-x
151 -y+1/2,z+1/2,-x
152 -y+1/2,-z+1/2,x
153 z+1/2,x+1/2,y
```

```
154 z+1/2,-x+1/2,-y
155 -z+1/2,x+1/2,-y
156 -z+1/2,-x+1/2,y
157 -y+1/2,-x+1/2,-z
158 -y+1/2,x+1/2,z
159 y+1/2,-x+1/2,z
160 y+1/2,x+1/2,-z
161 -x+1/2,-z+1/2,-y
162 -x+1/2,z+1/2,y
163 x+1/2,-z+1/2,y
164 x+1/2,z+1/2,-y
165 -z+1/2,-y+1/2,-x
166 -z+1/2,y+1/2,x
167 z+1/2,-y+1/2,x
168 z+1/2,y+1/2,-x
169 -x+1/2,-y+1/2,-z
170 -x+1/2,y+1/2,z
171 x+1/2,-y+1/2,z
172 x+1/2,y+1/2,-z
173 -y+1/2,-z+1/2,-x
174 -y+1/2,z+1/2,x
175 y+1/2,-z+1/2,x
176 y+1/2,z+1/2,-x
177 -z+1/2,-x+1/2,-y
178 -z+1/2,x+1/2,y
179 z+1/2,-x+1/2,y
180 z+1/2,x+1/2,-y
181 y+1/2,x+1/2,z
182 y+1/2,-x+1/2,-z
183 -y+1/2,x+1/2,-z
184 -y+1/2,-x+1/2,z
185 x+1/2,z+1/2,y
186 x+1/2,-z+1/2,-y
187 -x+1/2,z+1/2,-y
188 -x+1/2,-z+1/2,y
189 z+1/2,y+1/2,x
190 z+1/2,-y+1/2,-x
191 -z+1/2,y+1/2,-x
192 -z+1/2,-y+1/2,x
loop_
_atom_site_label
_atom_site_type_symbol
_atom_site_symmetry_multiplicity
_atom_site_Wyckoff_label
_atom_site_fract_x
_atom_site_fract_y
_atom_site_fract_z
        _atom_site_occupancy
        _atom_site_fract_symmform
Ag1 Ag   4 a 0.00000 0.00000 0.00000 1.00000 0,0,0
F1  F    4 b 0.50000 0.50000 0.50000 1.00000 0,0,0

AgO--CuO-type.cif
_audit_creation_method FINDSYM
```

```
_cell_length_a      5.6187910000
_cell_length_b      3.5156470000
_cell_length_c      5.4269620000
_cell_angle_alpha  90.0000000000
_cell_angle_beta  105.8654250000
_cell_angle_gamma  90.0000000000
_cell_volume       103.1187783941
_symmetry_space_group_name_H-M "P 1 21/c 1"
_symmetry_Int_Tables_number 14
_space_group.reference_setting '014:-P 2ybc'
_space_group.transform_Pp_abc a,b,c;0,0,0
loop_
_space_group_symop_id
_space_group_symop_operation_xyz
1 x,y,z
2 -x,y+1/2,-z+1/2
3 -x,-y,-z
4 x,-y+1/2,z+1/2
loop_
_atom_site_label
_atom_site_type_symbol
_atom_site_symmetry_multiplicity
_atom_site_Wyckoff_label
_atom_site_fract_x
_atom_site_fract_y
_atom_site_fract_z
     _atom_site_occupancy
     _atom_site_fract_symmform
Ag1 Ag  2 a 0.00000 0.00000 0.00000 1.00000 0,0,0
Ag2 Ag  2 d 0.50000 0.00000 0.50000 1.00000 0,0,0
O1  O   4 e 0.29002 0.84122 0.72896 1.00000 Dx,Dy,Dz

alpha-F2.cif
_audit_creation_method FINDSYM
_cell_length_a      5.5012970000
_cell_length_b      3.2304240000
_cell_length_c      7.1471600553
_cell_angle_alpha  90.0000000000
_cell_angle_beta  108.6099373188
_cell_angle_gamma  90.0000000000
_cell_volume       120.3746398387
_symmetry_space_group_name_H-M "C 1 2/m 1"
_symmetry_Int_Tables_number 12
_space_group.reference_setting '012:-C 2y'
_space_group.transform_Pp_abc a,b,c;0,0,0
loop_
_space_group_symop_id
_space_group_symop_operation_xyz
1 x,y,z
2 -x,y,-z
3 -x,-y,-z
4 x,-y,z
5 x+1/2,y+1/2,z
6 -x+1/2,y+1/2,-z
```

```
7 -x+1/2,-y+1/2,-z
8 x+1/2,-y+1/2,z
loop_
_atom_site_label
_atom_site_type_symbol
_atom_site_symmetry_multiplicity
_atom_site_Wyckoff_label
_atom_site_fract_x
_atom_site_fract_y
_atom_site_fract_z
        _atom_site_occupancy
        _atom_site_fract_symmform
F1 F    4 i 0.08273 0.00000 0.09882 1.00000 Dx,0,Dz
F2 F    4 i 0.49921 0.00000 0.59865 1.00000 Dx,0,Dz

alpha-O2--AFM.cif
_audit_creation_method FINDSYM
_cell_length_a      3.5417520000
_cell_length_b      3.4401490000
_cell_length_c      4.2881025246
_cell_angle_alpha 90.0000000000
_cell_angle_beta   110.0779368054
_cell_angle_gamma 90.0000000000
_cell_volume        49.0716777848
_symmetry_space_group_name_H-M "C 1 2/m 1"
_symmetry_Int_Tables_number 12
_space_group.reference_setting '012:-C 2y'
_space_group.transform_Pp_abc a,b,c;0,0,0
loop_
_space_group_symop_id
_space_group_symop_operation_xyz
1 x,y,z
2 -x,y,-z
3 -x,-y,-z
4 x,-y,z
5 x+1/2,y+1/2,z
6 -x+1/2,y+1/2,-z
7 -x+1/2,-y+1/2,-z
8 x+1/2,-y+1/2,z
loop_
_atom_site_label
_atom_site_type_symbol
_atom_site_symmetry_multiplicity
_atom_site_Wyckoff_label
_atom_site_fract_x
_atom_site_fract_y
_atom_site_fract_z
        _atom_site_occupancy
        _atom_site_fract_symmform
O1 O    4 i -0.06296 0.00000 0.84805 1.00000 Dx,0,Dz
```

# References

[1] W. Grochala, R. Hoffmann, "Real and Hypothetical Intermediate-Valence Fluoride $Ag^{II}/Ag^{III}$ and $Ag^{II}/Ag^{I}$ Systems as Potential Superconductors", Angew. Chem. Int. Ed. Engl. 40, 2742–2781 (2001). https://doi.org/10.1002/1521-3773(20010803)40:15<2742::AID-ANIE2742>3.0.CO;2-X

[2] Z. Mazej, D. Kurzydłowski, W. Grochala, "Unique silver(II) fluorides: the emerging electronic and magnetic materials", in: Photonic and Electronic Properties of Fluoride Materials, Eds. A. Tressaud & K. Poeppelmeier, pp. 231-260, Elsevier (2016).

[3] W. Grochala, Z. Mazej, "Chemistry of Ag(II): a cornucopia of peculiarities", Phil. Trans. A 373, 20140179 (2015). https://doi.org/10.1098/rsta.2014.0179

[4] T. Jaroń, W. Grochala, "Prediction of Giant Antiferromagnetic Coupling in Exotic Fluorides of $Ag^{II}$", Phys. Stat. Sol. RRL 2, 71–73 (2008). https://doi.org/10.1002/pssr.200701286

[5] D. Kurzydłowski, W. Grochala, "Prediction of extremely strong antiferromagnetic superexchange in silver(II) fluorides: challenging the oxocuprates(II)", Angew. Chem. Int. Ed. Engl. 56, 10114–10117 (2017). https://doi.org/10.1002/anie.201700932

[6] J. Gawraczyński, et al., "Silver route to cuprate analogs", Proc. Nat. Acad. Sci. USA 116, 1495–1500 (2019). https://doi.org/10.1073/pnas.1812857116

[7] A. Jesih, K. Lutar, B. Žemva, B. Bachmann, St. Becker, B. G. Müller, R. Hoppe, "Einkristalluntersuchungen an $AgF_2$", Z. Anorg. Allg. Chem. 588, 77 – 83 (1990). https://doi.org/10.1002/zaac.19905880110

[8] P. Fischer, D. Schwarzenbach, H. M. Rietveld, "Crystal and magnetic structure of silver difluoride: I. Determination of the $AgF_2$ structure", J. Phys. Chem. Solids 32, 543–550 (1971). https://doi.org/10.1016/0022-3697(71)90003-5

[9] J. A. McMillan, "The crystalline structure of AgO", Acta Cryst. 7, 640–640 (1954). No doi.

[10] A. J. Salkind, W. C. Zeek, "The Structure of AgO", J. Electrochem. Soc. 106, 366–366 (1959). No doi

[11] J. A. McMillan, "Magnetic properties and crystalline structure of AgO", J. Inorg. Nucl. Chem. 13, 28–31 (1960). https://doi.org/10.1016/0022-1902(60)80231-X

[12] V. Scatturin, P. Bellon, A. J. Salkind, "Structure of silver oxide AgO by neutron diffraction", Ricerca Scientifica 30, 1034–1044 (1960). No doi.

[13] K. Yvon, A. Bezinge, P. Tissot, P. Fischer, "Structure and magnetic properties of tetragonal silver(I,III) oxide, AgO", J. Solid State Chem. 65, 225–230 (1986). https://doi.org/10.1016/0022-4596(86)90057-5

[14] S. Riedel, M. Kaupp, "The highest oxidation states of the transition metal elements", Coord. Chem. Rev. 253, 606–624 (2009). https://doi.org/10.1016/j.ccr.2008.07.014

[15] O. Clemens, P. R. Slater, "Topochemical modifications of mixed metal oxide compounds by low-temperature fluorination routes", Rev. Inorg. Chem. 33, 105–117 (2013). https://doi.org/10.1515/revic-2013-0002

[16] J. P. Allen, D. O. Scanlon, G. W. Watson, "Electronic structure of mixed-valence silver oxide AgO from hybrid density-functional theory", Phys. Rev. B 81, 1–4 (2010). https://doi.org/10.1103/PhysRevB.81.161103

[17] M. Derzsi, P. Piekarz, W. Grochala, "Structures of Late Transition Metal Monoxides from Jahn-Teller Instabilities in the Rock Salt Lattice", Phys. Rev. Lett. 113, 1–5 (2014). https://doi.org/10.1103/PhysRevLett.113.025505

[18] J. T. Wolan, G. B. Hoflund, "Surface characterization study of AgF and $AgF_2$ powders using XPS and ISS", Appl. Surf. Sci. 125, 251–258 (1998). https://doi.org/10.1016/S0169-4332(97)00498-4

[19] W. Grochala, R. G. Egdell, P. P. Edwards, Z. Mazej, B. Žemva, "On the Covalency of Silver–Fluorine Bonds in Compounds of Silver(I), Silver(II) and Silver(III)", ChemPhysChem 4, 997–1001 (2003). https://doi.org/10.1002/cphc.200300777

[20] J. F. Weaver, G. B. Hoflund, "Surface Characterization Study of the Thermal Decomposition of AgO", Chem. Mater. 6, 8519–8524 (1994). https://doi.org/10.1021/j100085a035

[21] A. Grzelak, T. Jaroń, Z. Mazej, T. Michałowski, P. Szarek, W. Grochala, "Anomalous Chemical Shifts in X-ray Photoelectron Spectra of Sulfur-Containing Compounds of Silver (I) and (II)", J. Electr. Spectr. Rel. Phenom. 202, 38–45 (2015). https://doi.org/10.1016/j.elspec.2015.02.013

[22] A. Grzelak, et al., " High pressure behavior of silver fluorides up to 40 GPa", Inorg. Chem. 56, 14651–14661 (2017). https://doi.org/10.1021/acs.inorgchem.7b02528

[23] A. Grzelak, et al., "Persistence of mixed valence in the high-pressure structure of silver (I,III) oxide AgO: a combined Raman, XRD and DFT study", Inorg. Chem. 56, 5804–5812 (2017). https://doi.org/10.1021/acs.inorgchem.7b00405

[24] M. B. Robin, P. Day, "Mixed Valence Chemistry-A Survey and Classification", Adv. Inorg. Chem. Radiochem. 10, 247–422 (1968). https://doi.org/10.1016/S0065-2792(08)60179-X

[25] J. Zaanen, G. A. Sawatzky, J. W. Allen, "Band gaps and electronic structure of transition-metal compounds", Phys. Rev. Lett. 55, 418–421 (1985). https://doi.org/10.1103/PhysRevLett.55.418

[26] D. C. Lonie, E. Zurek, "XtalOpt: An open-source evolutionary algorithm for crystal structure prediction", Comput. Phys. Commun. 182, 372–387 (2011). https://doi.org/10.1016/j.cpc.2010.07.048

[27] M. Derzsi, A. Grzelak, P. Kondratiuk, K. Tokar, W. Grochala, "Quest for Compounds at the Verge of Charge Transfer Instabilities: The Case of Silver(II) Chloride", Crystals 9, 423 (2019). https://doi.org/10.3390/cryst9080423

[28] A. I. Liechtenstein, V. I. Anisimov, J. Zaanen, "Density-functional theory and strong interactions: Orbital ordering in Mott-Hubbard insulators", Phys. Rev. B, 52, R5467–R5470 (1995). https://doi.org/10.1103/PhysRevB.52.R5467

[29] J. P. Perdew, A. Ruzsinszky, G. I. Csonka, O. A. Vydrov, G. E. Scuseria, L. A. Constantin, X. Zhou, K. Burke, "Restoring the Density-Gradient Expansion for Exchange in Solids and Surfaces", Phys. Rev. Lett. 100, 136406 (2008). https://doi.org/10.1103/PhysRevLett.100.136406

[30] D. Kasinathan, K. Koepernik, U. Nitzsche, H. Rosner, "Ferromagnetism Induced by Orbital Order in the Charge-Transfer Insulator $Cs_2AgF_4$: An Electronic Structure Study", Phys. Rev. Lett. 99, 247210 (2007). https://doi.org/10.1103/PhysRevLett.99.247210

[31] P. E. Blöchl, "Projector augmented-wave method", Phys. Rev. B 50, 17953–17979 (1994). https://doi.org/10.1103/PhysRevB.50.17953

[32] G. Kresse, J. Hafner, "Ab initio molecular-dynamics simulation of the liquid-metal–amorphous-semiconductor transition in germanium", Phys. Rev. B, 49, 14251–14269 (1994). https://doi.org/10.1103/PhysRevB.49.14251

[33] "The PAW potentials for d-elements," can be found under https://cms.mpi.univie.ac.at/vasp/vasp/_elements.html, 2020.

[34] B. Darriet, J. Galy, "Synthèse et structure cristalline du bis[difluorooxostannate(II)] d'étain(II), $(Sn_2O_2F_4)Sn_2$", Acta Cryst. B 33, 1489–1492 (1977). https://doi.org/10.1107/S0567740877006359

[35] A. Bystroem, "The structure of the fluorides and oxifluoride of bivalent lead", Arkiv foer Kemi, Mineralogi och Geologi 24, 33–33 (1947). No doi.

[36] B. Aurivillius, "X-ray studies of lead oxide fluoride and related compounds", Chemica Scripta 10, 156–158 (1976). No doi.

[37] L. Noodleman, "Valence bond description of antiferromagnetic coupling in transition metal dimers", J. Chem. Phys. 74, 5737–5743 (1981). https://doi.org/10.1063/1.440939

[38] D. Dai, M.-H. Whangbo, "Spin-Hamiltonian and density functional theory descriptions of spin exchange interactions", J. Chem. Phys. 114, 2887–2893 (2001). https://doi.org/10.1063/1.1342758

[39] F. Illas, I. D. P. R. Moreira, C. de Graaf, V. Barone, "Spin Symmetry Requirements in Density Functional Theory: The Proper Way to Predict Magnetic Coupling Constants in Molecules and Solids", Theor. Chem. Acc. 104, 265–272 (2000). https://doi.org/10.1007/s00214-006-0104-6

[40] K. Momma, F. Izumi, "VESTA 3 for three-dimensional visualization of crystal, volumetric and morphology data", J. Appl. Cryst. 44, 1272–1276 (2011). https://doi.org/10.1107/S0021889811038970

[41] L. Q. Hatcher, K. D. Karlin, "Oxidant types in copper-dioxygen chemistry: the ligand coordination defines the Cu(n)-O2 structure and subsequent reactivity", J. Biol. Inorg. Chem. 9, 669–683 (2004). https://doi.org/10.1007/s00775-004-0578-4

[42] C. Würtele, E. Gaoutchenova, K. Harms, M. C. Holthausen, J. Sundermeyer, S. Schindler, "Crystallographic Characterization of a Synthetic 1:1 End-On Copper Dioxygen Adduct Complex", Angew. Chem. Int. Ed. Engl. 45, 3867–3869 (2006). https://doi.org/10.1002/anie.200600351

[43] P. J. Malinowski, M. Derzsi, B. Gaweł, W. Łasocha, Z. Jagličić, Z. Mazej, W. Grochala, "$Ag^{II}SO_4$: A Genuine Sulfate of Divalent Silver with Anomalously Strong One-Dimensional Antiferromagnetic Interactions", Angew. Chem. Int. Ed. Engl. 49, 1683–1686 (2010). https://doi.org/10.1002/anie.200906863

[44] P. J. Malinowski, M. Derzsi, Z. Mazej, Z. Jagličić, P. J. Leszczyński, T. Michałowski, W. Grochala, "Silver(II) Fluorosulfate: A Thermally Fragile Ferromagnetic Derivative of Divalent Silver in an Oxa-Ligand Environment", Eur. J. Inorg. Chem. 2499–2507 (2011). https://doi.org/10.1002/ejic.201100077

[45] B. Standke, M. Jansen, "$Ag_3O_4$, the First Silver(II,III) Oxide", Angew. Chem. Int. Ed. Engl. 25, 77–78 (1986). https://doi.org/10.1002/anie.198600771

[46] B. Stehlik, P. Weidenthaler, J. Vlach, "Kristallstruktur von silber(III)-oxyd", Coll. Czech. Chem. Commun. 24, 1581–1588 (1959). https://doi.org/10.1135/cccc19591581

[47] B. Standke, M. Jansen, "Darstellung und Kristallstruktur von $Ag_2O_3$", Z. Anorg. Allg. Chem. 535, 39–46 (1986). https://doi.org/10.1002/zaac.19865350406

[48] D. Kurzydłowski, Z. Mazej, Z. Jagličić, Y. Filinchuk, W. Grochala, "Structural transition and unusually strong 1D antiferromagnetic superexchange coupling in perovskite $KAgF_3$", Chem. Commun. 49, 6262–6264 (2013). https://doi.org/10.1039/C3CC41521J

[49] W. Grochala, "Silverland: the realm of compounds of divalent silver – and why they are interesting", J. Supercond. Novel Magnet. 31, 737–752 (2018). https://doi.org/10.1007/s10948-017-4326-8

[50] P. Połczyński, R. Jurczakowski, W. Grochala, "Stabilization and strong oxidizing properties of Ag(II) in a fluorine-free solvent", Chem. Commun. 49, 7480–7482 (2013). https://doi.org/10.1039/C3CC43072C

[51] R. J. Meier, R. B. Helmholdt, "Neutron-diffraction study of α- and β-oxygen", Phys. Rev. B 29, 1387–1393 (1984). https://doi.org/10.1103/PhysRevB.29.1387

[52] L. Meyer, C. S. Barrett, S. C. Greer, "Crystal Structure of α-Fluorine", J. Chem. Phys. 49, 1902–1907 (1968). https://doi.org/10.1063/1.1670323